\documentclass[usenatbib,usedcolumn,usegraphicx]{mn2e}

\newcommand{\tabc}[1]{\multicolumn{1}{c}{#1}}
\newcommand{\sub}[1]{_{\rm {#1}}}
\newcommand{\mach}{{\cal M}}
\newcommand{\fJ}{f\sub{J}}
\newcommand{\cs}{c\sub{s}}
\newcommand{\tcoll}{t\sub{coll}}
\newcommand{\tcross}{t\sub{cross}}

\begin{document}

\title{Hydrodynamical simulations of a cloud of interacting gas fragments}
\author[D.~Gittins, C.~Clarke and M.~Bate]
{D.~M.~Gittins,$^1$\thanks{Email: dgittins@ast.cam.ac.uk} C.~J.~Clarke$^1$ and M.~R.~Bate$^2$\\
$^1$Institute of Astronomy, Madingley Road, Cambridge CB3 0HA\\
$^2$School of Physics, University of Exeter, Stocker Road, Exeter EX4 4QL}

\maketitle

\begin{abstract}
We present a full hydrodynamical investigation of a system of self-contained, interacting
gas `cloudlets'.  Calculations are performed using SPH, and cloudlets are allowed
to collide, dissipate kinetic energy, merge and undergo local gravitational collapse.
The numerical feasibility of such a technique is explored, and the resolution requirements examined.
We examine the effect of the collision time and the velocity field on the timescale for
evolution of the system, and show that almost all star formation will be confined to dense regions
much smaller than the entire cloud.  We also investigate the possibility, discussed by
various authors, that such a system could lead to a power-law IMF by a process of
repeated cloudlet coagulation, finding that in fact the inter-cloudlet collisions
occur at too high a Mach number for merging to play an important part.
\end{abstract}


\section{Introduction}
\label{intro}

Star forming regions are often interpreted in a `cloud-clump-core' picture,
in which an overall system (the cloud) consists of an ensemble of
discrete regions of gas (clumps), which move about with supersonic
velocities through some confining medium.  The collisions and mergers
of these clumps lead to regions of gravitational collapse (cores) and
eventually to star formation.  This idea is largely an attempt to
characterise the observed structure of molecular cloud complexes
(see for example the review by \citealt*{wbm}),
but is likely to be applicable in many situations.
Various authors such as \citet{kwan} and \citet{st}
have cited coagulation processes as an origin for the form of the IMF.
For example, \citet{ML} have argued
that in protoglobular clusters, gas is fragmented into sub-Jeans mass
clumps by a variety of hydrodynamical instabilities, and that eventual
star formation ensues when enough clumps have merged to form Jeans
unstable objects.
A similar process may be involved in the assembly of
Giant Molecular Clouds (GMCs) in the spiral arms of disc galaxies \citep{pal}, 
since the diffuse ISM that is gathered together in galactic shocks is
inferred to be highly clumped on small scales \citep{koyama}.
It is the interactions of these clumps, or `cloudlets', that controls the
formation of stars, and understanding the properties of such systems is important
in order to shed light on questions such as the timescales for star
formation and the origin of the IMF.

Until recently, a full treatment of such a system was
beyond the computing resources available.
Previous authors have
examined numerically the interaction of gas cloudlets, but the majority
of these studies have been limited to the collision of two objects.
Early work (\citealt{coll1}; \citealt{coll2}) was severely limited
by the resolution available.  \citet{coll3} carried out a thorough
examination of the possible outcomes of a two-body collision.  Later
studies (\citealt{coll4}; \citealt{coll5}; \citealt{coll6}; \citealt{coll7})
all examine in detail the structure produced in one collision between two
large bodies of gas.

None of these numerical investigations concern a system of many
interacting bodies.  One attempt to do so \citep{monaghan88} examined
a group of 48 cloudlets using SPH, but was limited to a very low resolution.
Instead,
authors have overcome the computing limitations by using
adapted N-body methods, in which the results of cloudlet collisions are
determined according to prescribed rules whenever two cloudlets approach each other
(\citealt{SP}; \citealt{ML}; \citealt*{murray99}).
A full hydrodynamical treatment of such a system using Smoothed Particle Hydrodynamics
is only now becoming computationally feasible, and it is
to this problem that we devote this paper.

Our aim is not to present a realistic model of any astronomical object,
but rather to investigate the numerical modelling of such ensembles.
We will examine the feasibility of such simulations, with particular
regard to the numerical resolution required, and discuss the limitations
of this approach.  We will also explore the general properties of
such systems, with particular regard to the energy dissipation timescale,
the distribution of protostars and the r\^{o}le of coagulation in the
evolution of the system.  In this way we will be able to reassess earlier
investigations such as that of \citet{ML}, who found that successive
clump-clump collisions built up a spectrum of resulting `stellar masses'
that was broadly compatible with a power law IMF.

 In addition to the abstract questions described above, this paper  also
lays the ground work for a series of papers investigating the formation
of GMCs from a shocked, clumpy ISM in the spiral arms of disc galaxies.
A key feature of these latter simulations involves the behaviour of
colliding clumps in the complex velocity field generated in spiral shocks,
and thus it is of considerable importance for these more realistic simulations
that the generic properties of coalescing clump ensembles, and the
numerical requirements for such calculations,  are well
understood.

The structure of the paper is as follows: in section \ref{numerical} we set
out the numerical method and the results of  a suite of two-body clump-clump
encounters. The aim here is to define  regions of parameter space
associated with particular outcomes (merger, collapse, or clump destruction)
and to define the number of particles per clump that is required
in order to achieve numerical convergence of expected outcomes. Section 
\ref{parameters} contains a detailed description of the system investigated
and outlines the key parameters.
In section \ref{results}  we describe generic properties of the cloud evolution and focus on the
following aspects:  the relationship between the energy dissipation
timescale  and the nominal two-body
collision timescale  (thus re-evaluating the claims made by \citet{SP}
in this regard), the possibility of such a system leading to
a distributed cluster of protostars by collisionally-induced local
gravitational collapse, and the process of building up a realistic
IMF/clump mass spectrum by this process.  Section \ref{conclusion} summarises our chief
conclusions.

\section{Numerical technique}
\label{numerical}

\subsection{Hydrodynamical code}
\label{code}

The calculation presented here was performed using a three-dimensional, 
smoothed particle hydrodynamics (SPH) code.  The SPH code is 
based on a version originally developed by Benz 
(\citealt{benz90}; \citealt{benz+90}).
The smoothing lengths of particles are variable in 
time and space, subject to the constraint that the number 
of neighbours for each particle must remain approximately 
constant at $N\sub{neigh}=50$.  The SPH equations are 
integrated using a second-order Runge-Kutta-Fehlberg 
integrator with individual time steps for each particle 
\citep*{bate95}.
Gravitational forces between particles and a particle's 
nearest neighbours are calculated using a binary tree.  
We use the standard form of artificial viscosity 
(\citealt{monaghan83}; \citealt{monaghan92}) with strength 
parameters $\alpha\sub{v}=1$ and $\beta\sub{v}=2$.
Further details can be found in \citet{bate95}.
The code has been parallelised by M.\ Bate using OpenMP.

The code makes use of two common modifications.  The first concerns the
fate of a region of self-gravitating material that begins to collapse.
Unchecked, the material will fall inward towards infinite density, and
as it does so the CPU time required to advance the gas rises without
limit.  This problem can be overcome by the use of `sink
particles' (see \citet{bate95} for more details).
In this technique, when an SPH particle passes a critical density
($\rho\sub{crit} = 10^6$ times the initial mean density),
a sink particle is formed by 
replacing the SPH gas particles contained within
$r\sub{acc} = 0.1 r\sub{cloudlet}$
of the densest gas particle in a pressure-supported fragment
by a point mass with the same mass and momentum.  Any gas that
later falls within this radius is accreted by the point mass
if it is bound and its specific angular momentum is less than
that required to form a circular orbit at radius $r\sub{acc}$
from the sink particle.
Sink particles interact with the gas only via gravity and accretion.
Sink particle mergers are not allowed.
Each sink particle therefore represents an area of the gas which has collapsed
under its own gravity.  The further fate of such a region is clearly not
resolved by the calculation.  These sink particles will be
referred to as `protostars' in the remainder of the paper.

The second modification is the inclusion of a confining external pressure.  The
cloudlets in our simulations are assumed to be moving through a thin confining medium,
consisting of a hot gas that exerts sufficient pressure to keep the undisturbed
cloudlets confined.  The cloudlets do not otherwise interact with the confining medium,
and exchange no momentum with it.
The
pressure of this external medium can be easily included in SPH by introducing a `constant
pressure' technique.  When the pressure force between two particles is calculated,
the pressure of each particle $p$ is replaced by a modified pressure $p^\prime = p - p\sub{ext}$,
where $p\sub{ext}$ is the constant external pressure, fixed at the start of the calculation.

The calculations presented here
used between $10^5$ and $10^6$ SPH particles.  The largest of these require
significant computing power to run in a practical length of time, and the UKAFF
supercomputing facility in Leicester was used for this purpose.

\subsection{Resolution requirements}
\label{resolution}

The resolution requirements must be carefully determined to ensure that
the calculation is always resolved, whilst keeping the total number of
SPH particles to a minimum.  This allows the calculations to be completed
in a reasonable length of time while following the evolution of the system
for as long as possible.

\citet{bate97} determined that the basic requirement for an SPH calculation
of this type is to ensure that the number
of SPH particles {\it per Jean's mass} remains above $2 N\sub{neigh}$ particles.  In the
simulations presented here, the cloudlets will collide with each other, and in the
colliding region an isothermal shock layer will be formed.
In such shocks, the density is increased by a factor equal to the square of
the Mach number $\mach$.  The Jean's mass
in such layers (and thus the number of particles per Jean's mass) will consequently fall
by a factor $\mach$.  In this way, given the expected Mach number of collisions,
a sufficient number of particles can be chosen to ensure that the resolution requirement
will be met in the isothermal shock layers.

In addition, two more tests are required to demostrate that the appropriate hydrodynamical
processes are being correctly simulated -- firstly, the kinetic energy dissipation in
a collision between two cloudlets, and secondly, the conditions under which such a collision
can lead to a collapse or a merger.  These two issues are discussed in the following sections.

\subsubsection{Kinetic energy dissipation in cloudlet collisions}

In a collision between two cloudlets, kinetic energy will be dissipated.  Since
this is the dominant process in the evolution of the systems we investigate, it is
important to check that this loss of energy is followed sufficiently accurately
at the numerical resolution we have used in our calculations.

We checked this requirement by performing a single two-body encounter with a range
of nine resolutions, from 50 up to $5\times 10^5$ SPH particles per cloudlet.
The calculation follows the collision of two spherical cloudlets of radius 1 at Mach 4,
at an impact parameter $b/r = 1$
(i.e.\ the separation perpendicular to the direction of travel is equal to
the radius of one sphere).    The calculation is performed in dimensionless units
such that the gravitational constant $G$, the sound speed of the gas $\cs$ and
the mass of each cloudlet are all equal to 1.  The cloudlets are Bonnor-Ebert
spheres with their boundary at $\xi \simeq 2.08$ (using the terminology of \citealt{bonnor56}),
corresponding to a density contrast
of $\rho\sub{centre} / \rho\sub{outer} \simeq 1.8$.
The external pressure is chosen to confine the spheres at the appropriate radius.
Under these conditions each sphere contains approximately $0.25$ Jean's masses.
The cloudlets are given
initial velocities of $v=2$ toward each other. At time $t=0$ the separation along the direction of
travel is 4 radii, such that if the cloudlets did not interact, their centres would pass each
other at time $t=1$.

\begin{figure}
\includegraphics[width=0.23\textwidth]{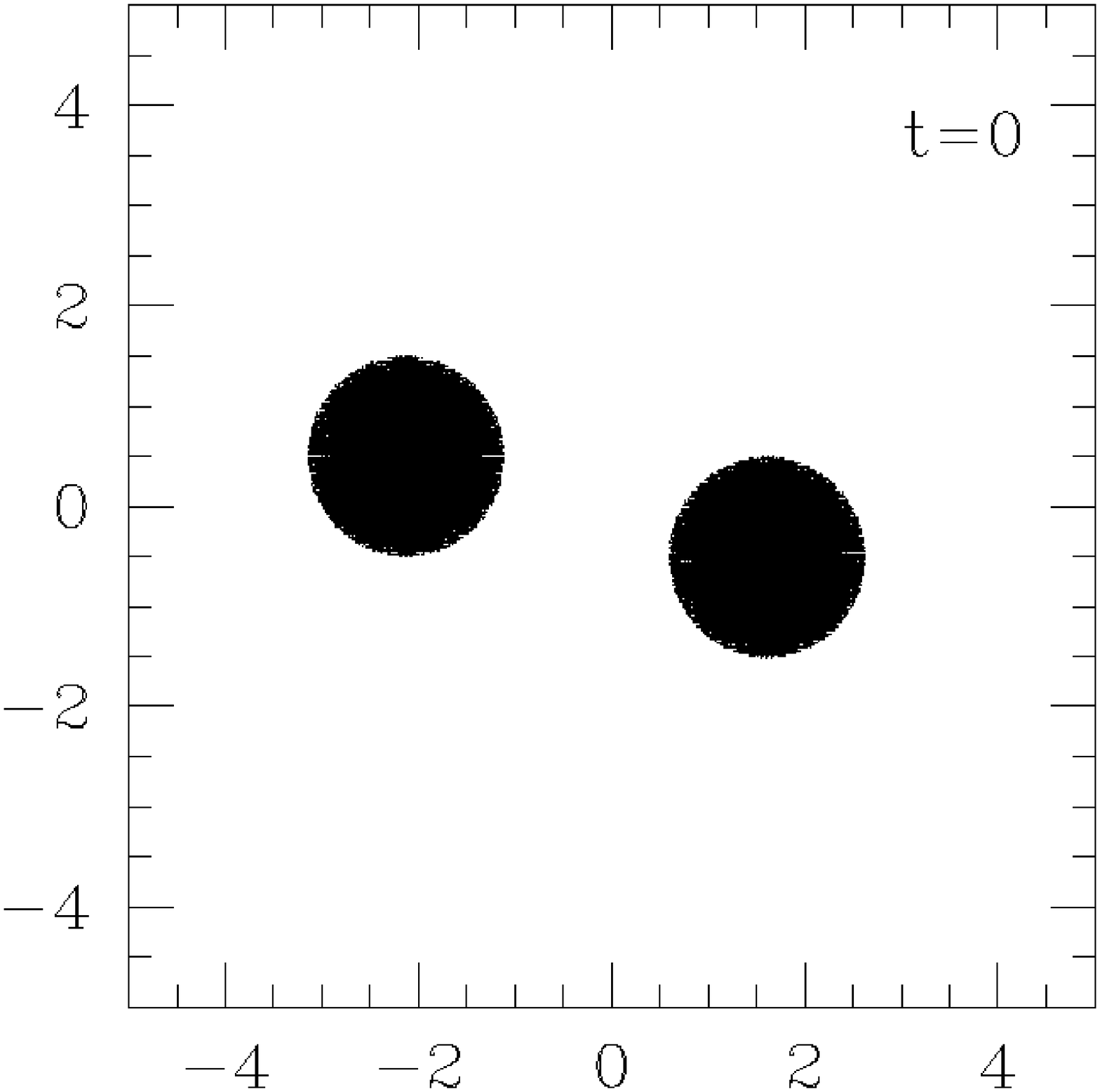}
\includegraphics[width=0.23\textwidth]{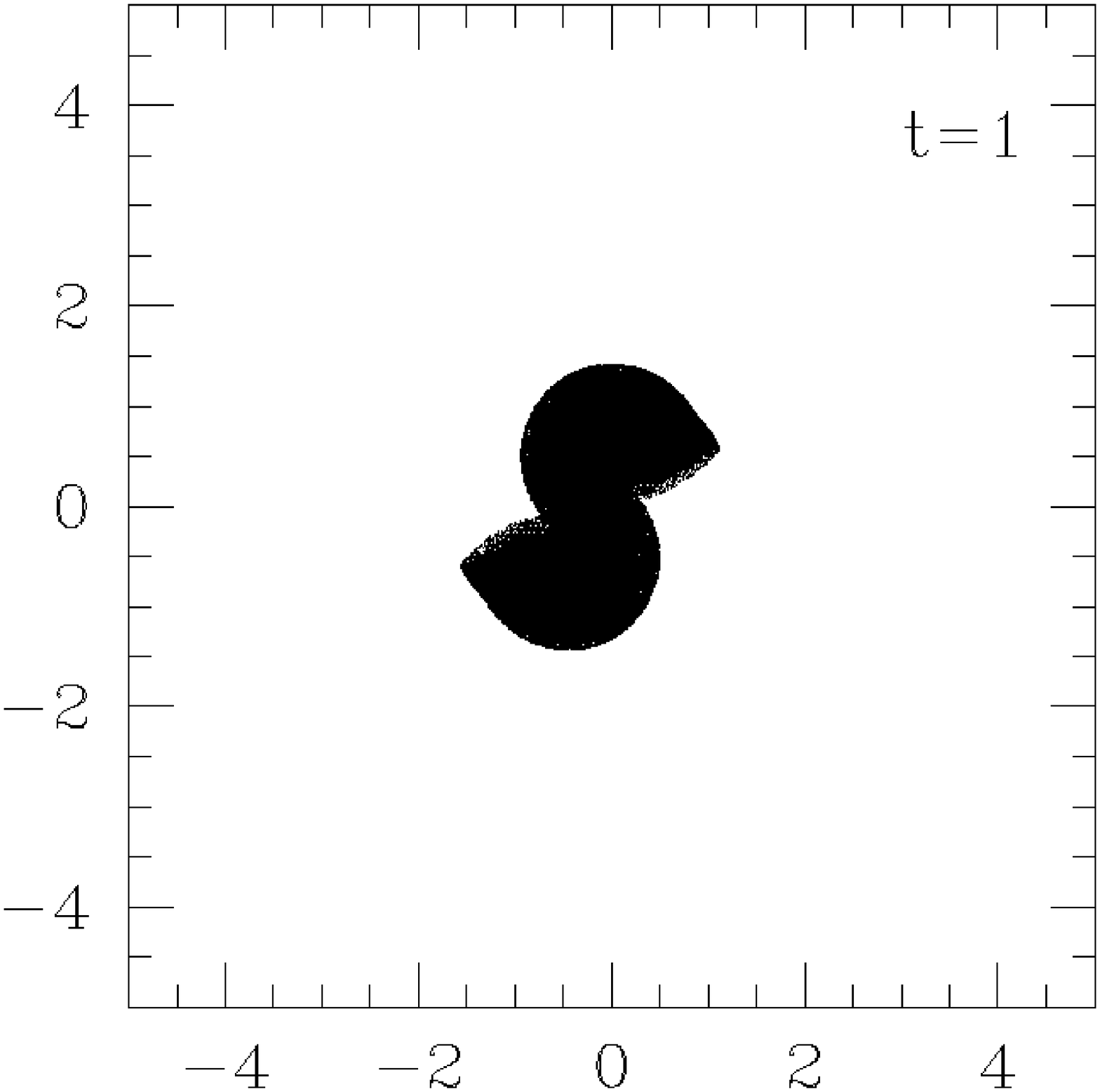}
\includegraphics[width=0.23\textwidth]{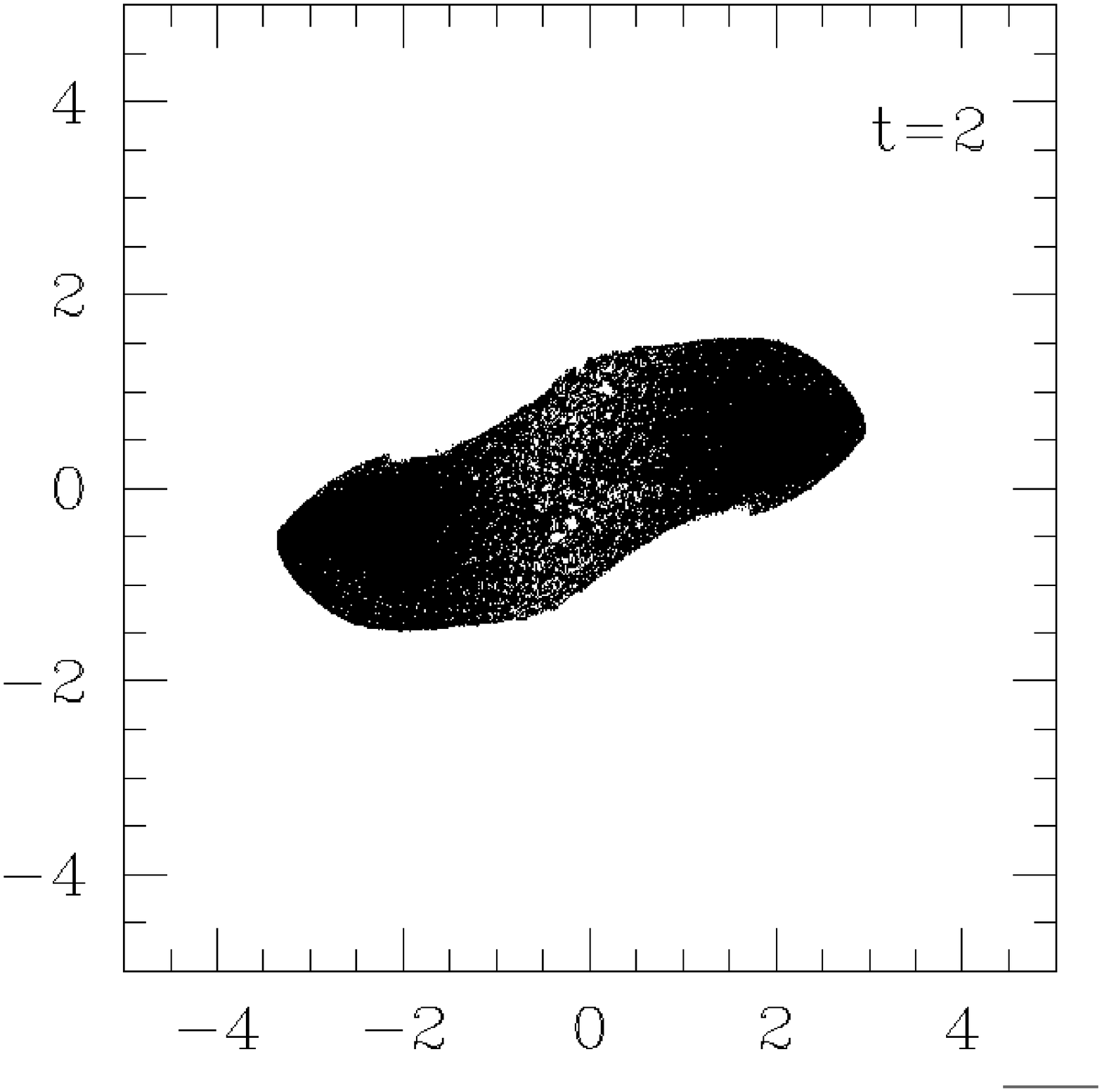}
\includegraphics[width=0.23\textwidth]{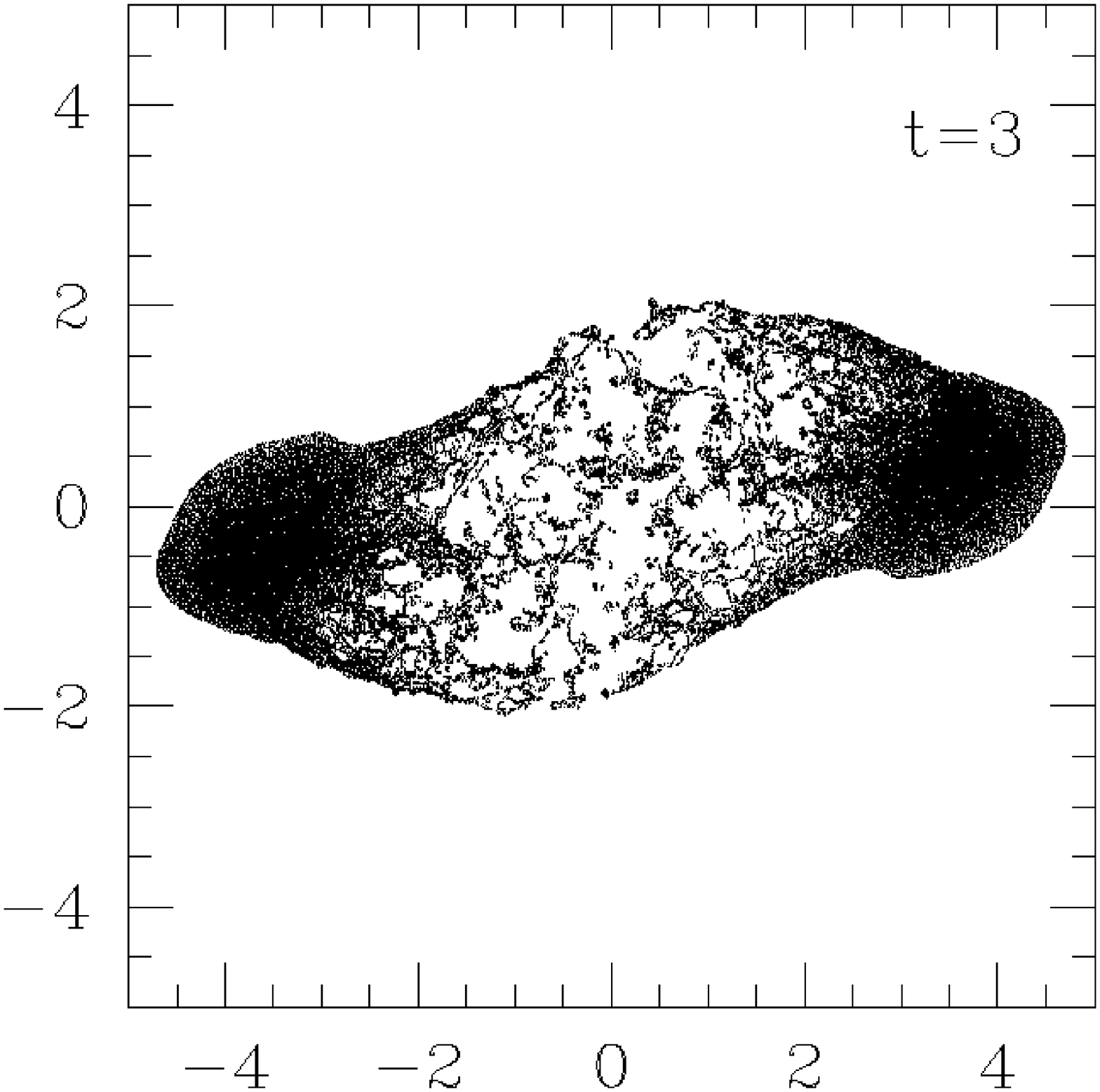}
\includegraphics[width=0.23\textwidth]{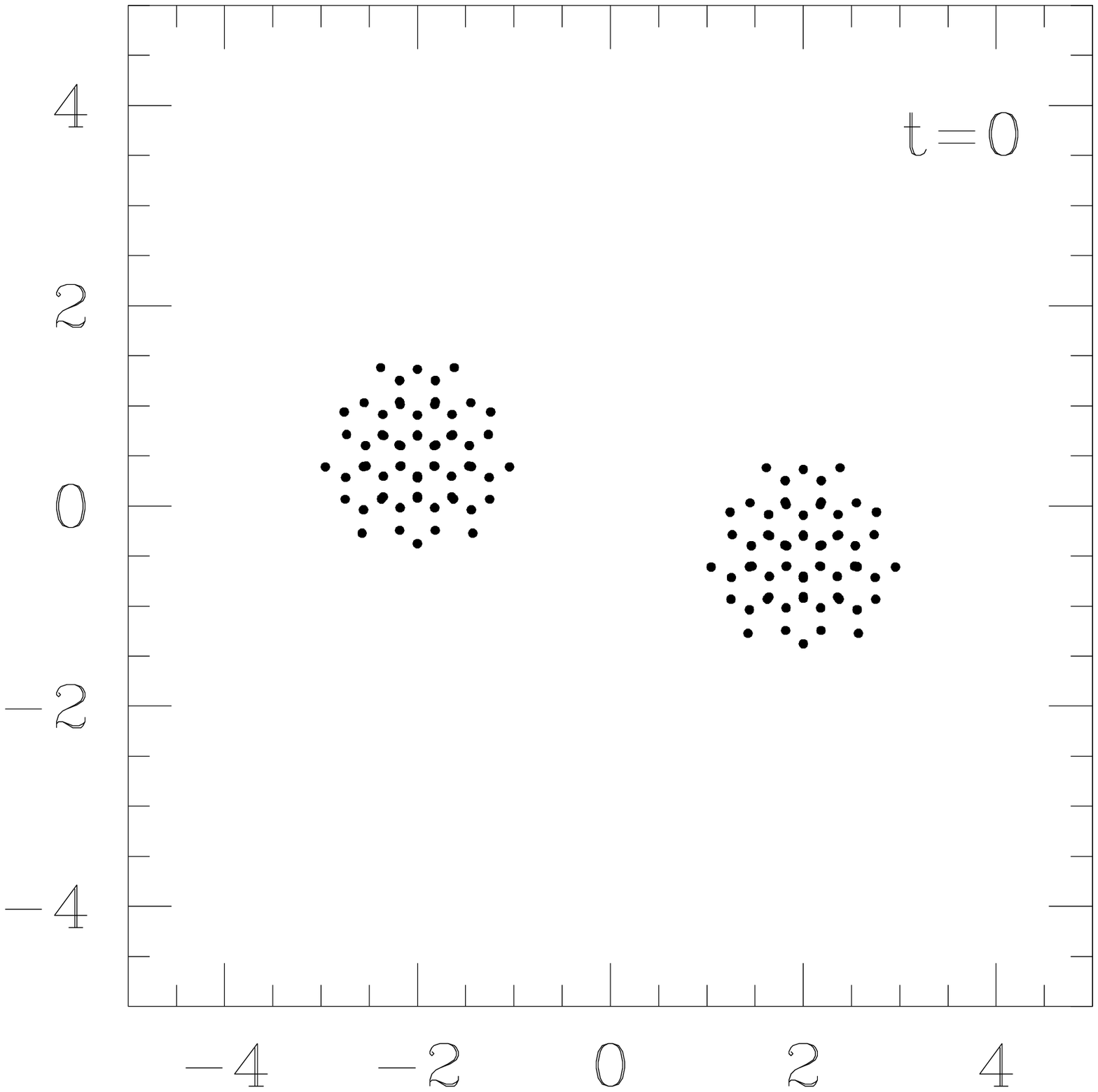}
\includegraphics[width=0.23\textwidth]{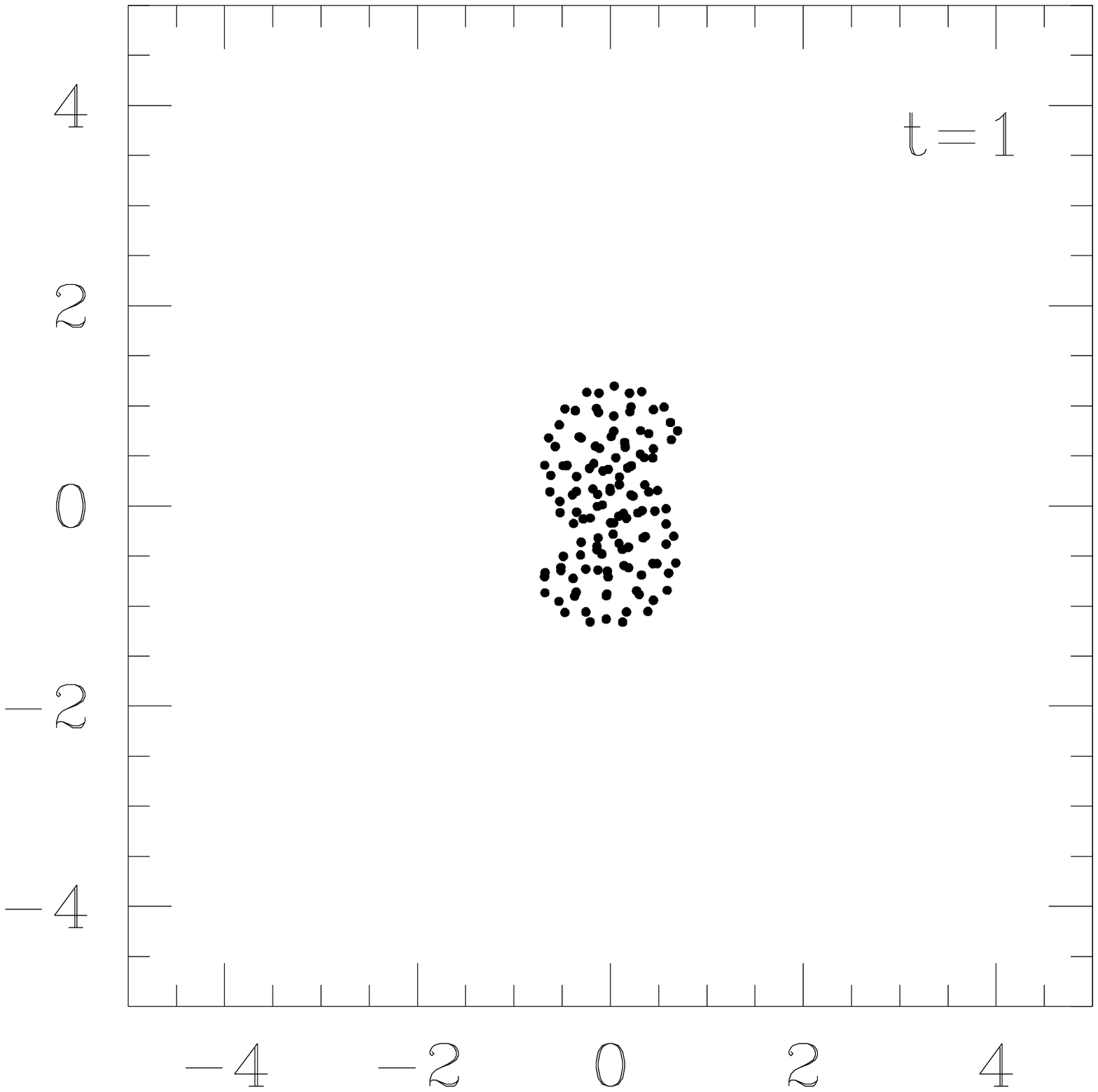}
\includegraphics[width=0.23\textwidth]{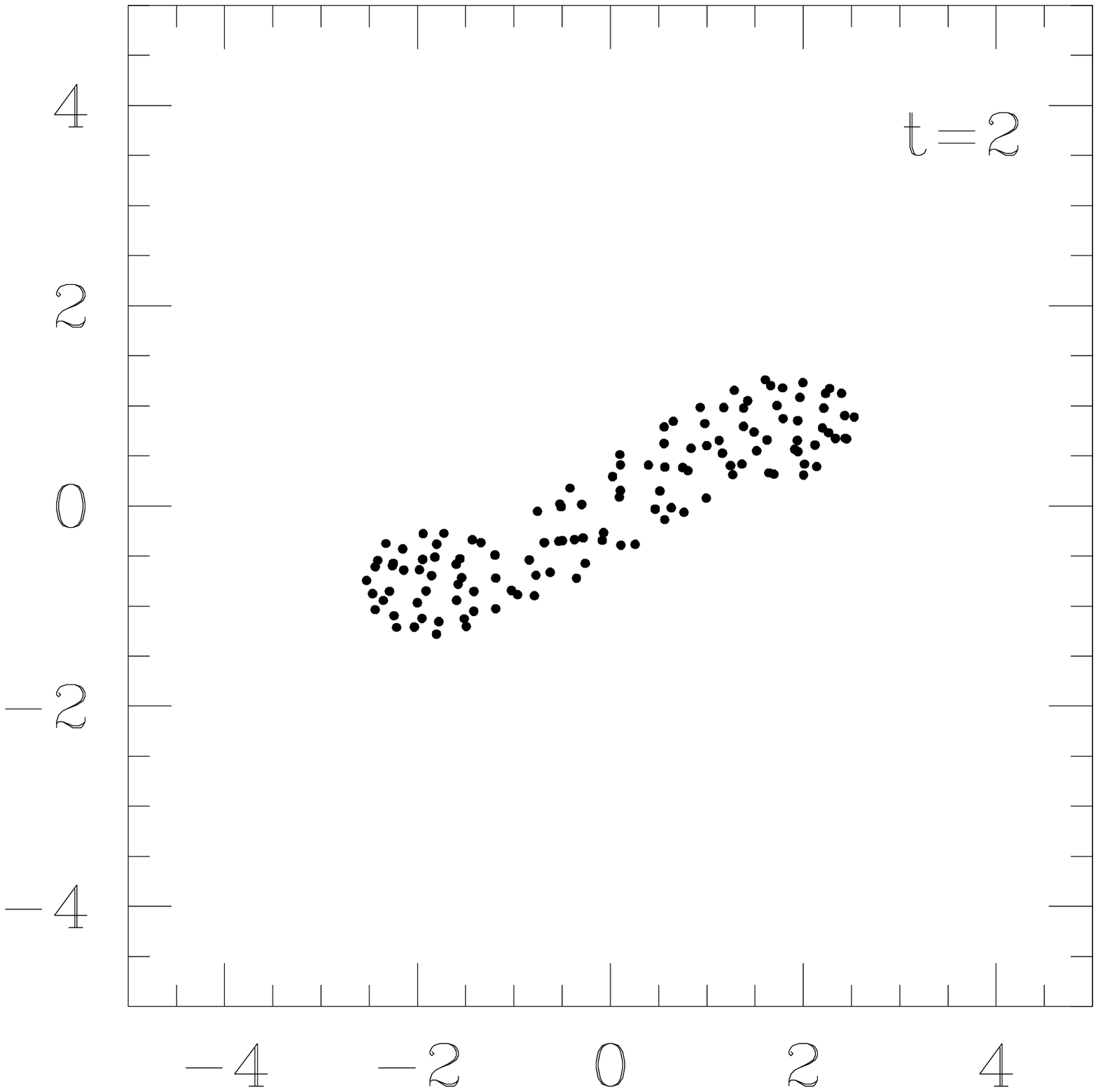}
\includegraphics[width=0.23\textwidth]{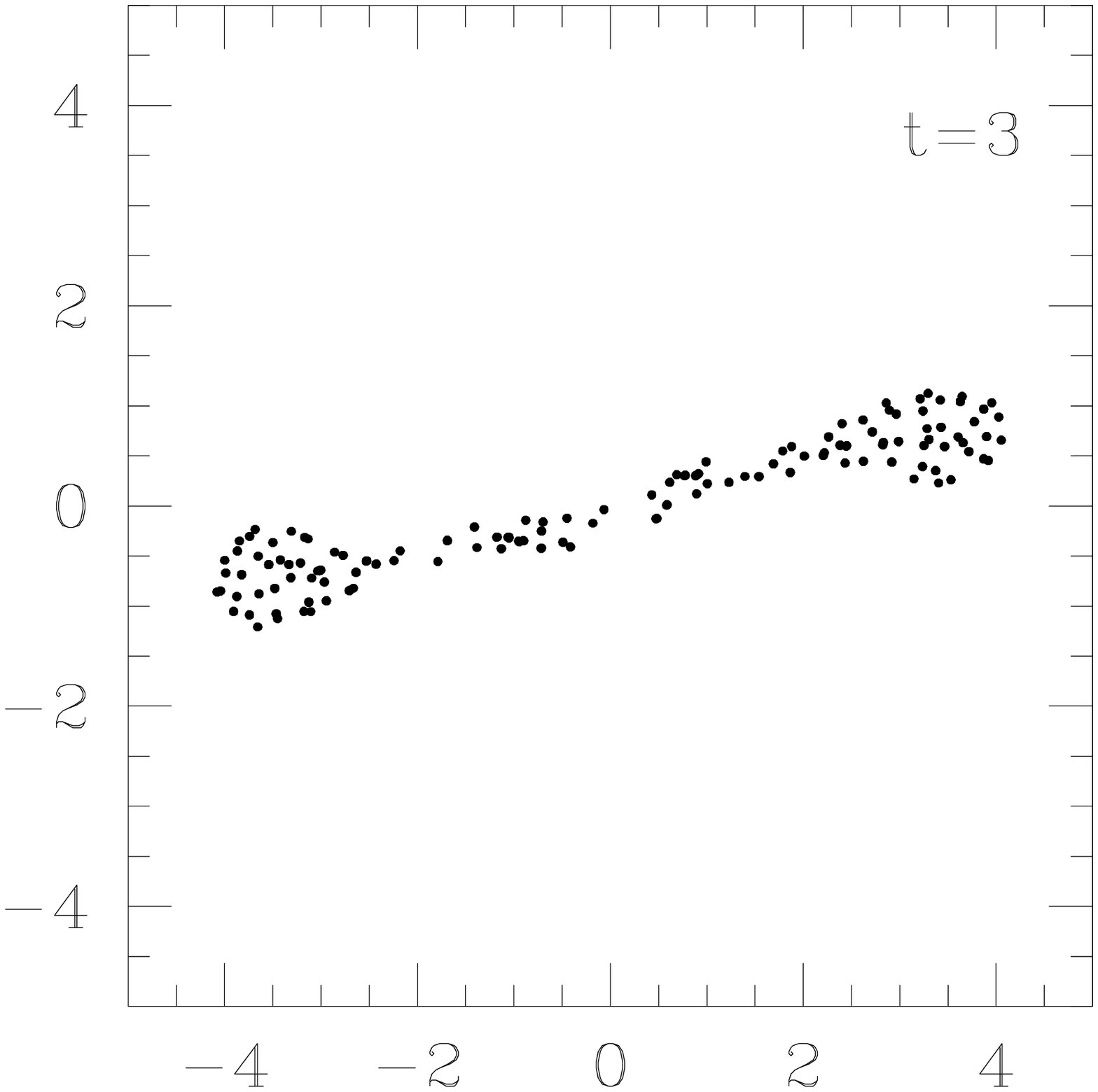}
\caption{Progress of the two-cloudlet collision used to test the kinetic energy dissipation in
collisions.  Two cloudlets of isothermal gas are collided at Mach 4.  The cloudlets are
Bonnor-Ebert spheres with $\xi \simeq 2.08$.
The initial separation along the direction of travel was 4 radii, and perpendicular to the direction of
travel was 1 radius.  The upper four panels show the calculation using 50,000 particles per cloudlet;
the lower four, using 100 particles per cloudlet.}
\label{twobodypic}
\end{figure}

\begin{figure}
\includegraphics[width=0.49\textwidth]{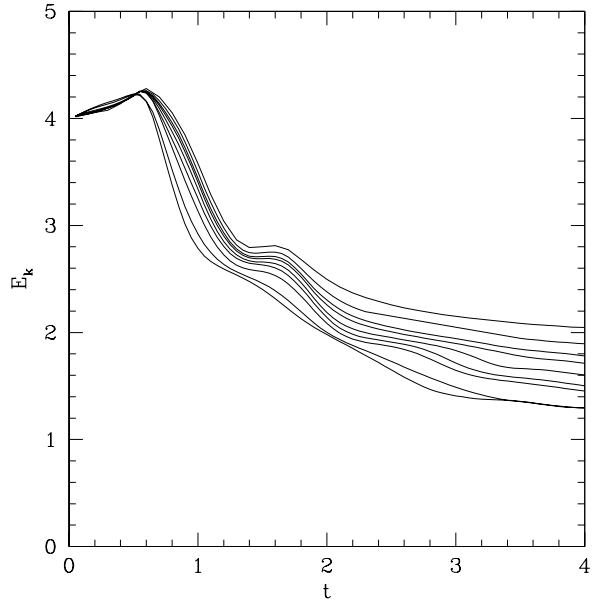}
\caption{Kinetic energy vs.\ time in the collision shown in figure \ref{twobodypic}.
The units are dimensionless (the gravitational constant,
sound speed, initial radii and mass of each sphere are 1).  In the absence of
any interaction, the centres of the cloudlets would pass each other at time $t=1$.  Nine different resolutions
are shown, with approximately the following numbers of SPH particles per cloudlet: (top to bottom)
500,000, 50,000, 15,000, 7,500, 3000, 1000, 500, 100 and 50.}
\label{twobodydiss}
\end{figure}

The progress of the collision is shown in figure \ref{twobodypic}.
Initially, the kinetic energy increases as the cloudlets
attract each other.  When they hit, a shock layer forms.
The pressure gradient tends to push particles away from the direction
of travel, and within the shock layer, much of the material
from one cloudlet avoids being reduced to zero velocity by
`slipping past' material from the other.
As the cloudlets separate, material is spread out between them.
As this gas expands, its reduced pressure will create a further deceleration on the
trailing edges of the initially non-overlapping regions of the cloudlets, dissipating
further kinetic energy.  This expanding region develops a complex structure
at high numerical resolution, which is unresolved when fewer particles are used.
The surviving portions of each cloudlet escape with a reduced velocity.
The total kinetic energy of the gas is displayed in figure \ref{twobodydiss}.

\begin{figure}
\includegraphics[width=0.49\textwidth]{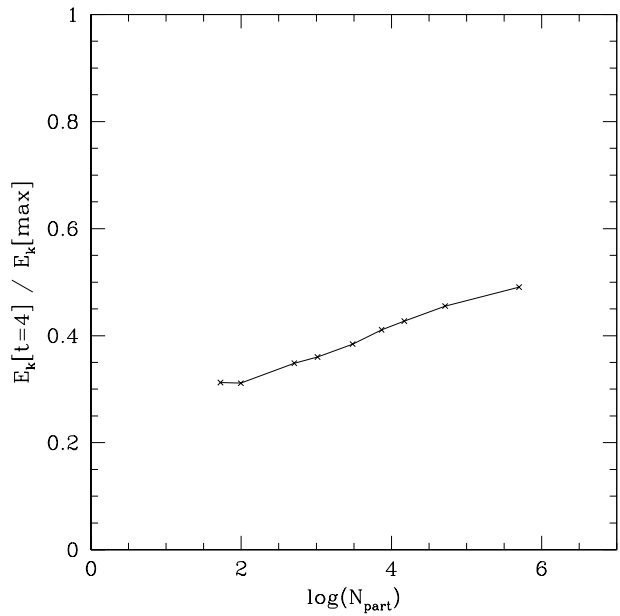}
\caption{Kinetic energy at time $t=4$ in the two-cloudlet collision shown in
figure \ref{twobodypic}, plotted against $\log(N)$ where $N$ is the number
of SPH particles per cloudlet.}
\label{twobodylogplot}
\end{figure}

As is clear from figure \ref{twobodydiss}, the greater the numerical resolution, the
less kinetic energy is dissipated in the collision.  This is to be expected, since
at lower resolution the smoothing lengths of the particles will be necessarily 
larger, and consequently, any dissipative region will be `smoothed' over a larger volume, and
hence a larger proportion of the mass.  In figure \ref{twobodylogplot}, the
kinetic energy of the system at time $t=4$ is plotted against the logarithm of
the number of particles.

It is somewhat difficult to assess the `realism' of this result.  A theoretical
value for the proportion of kinetic energy dissipated is not readily
obtainable.  Various previous authors have examined similar collisions, but few have
considered the energy dissipated; where such a value has been quoted, the effect of altering
the numerical resolution has not been examined.  It would appear from figure \ref{twobodylogplot}
that there may be a turnover around $N\sub{part} = 10^5$, implying
that the calculation would converge on some value at ever increasing
resolution.  It is clear that using only $100$ particles per cloudlet will
overestimate the amount of energy dissipated in each collision, but the
magnitude of this overestimation is not clear.  We conclude that
this is a fundamental limitation of attempting to model molecular
clouds using this technique, and that one should therefore avoid
relying on the detailed  numerical value of the kinetic energy in such systems,
exploring instead the overall evolution and structure of the cloud.

\subsubsection{Merging and collapse of colliding cloudlets}
\label{mergers}

The general problem of the collision of two clouds of gas has been examined by
many authors, as discussed in section \ref{intro}.  The range of conditions that can be simulated
is very large, and the range of outcomes is equally great (see for example \citealt{coll3}).
Any analysis must be restricted to a subset of the parameter space.
We intend here to examine the effect of varying numerical resolution on the outcome of a
simple two-body collision, of the type that dominates the calculations presented
later in this paper --- namely, the collision of two identical Bonnor-Ebert
spheres.

There are many possible outcomes from the collision of two such cloudlets.  Work on
simulating such encounters shows that these outcomes
can be divided into three categories: a) a collapsing object, in which the central region
becomes gravitationally unstable,
b) a merger, in which the end result is one large, stable cloudlet containing all the mass
of the two progenitors, or c) a `splash', in which instance the overlapping portion
of the gas tends to be ripped from the cloudlets and splashed into the region between
them, and two smaller regions of gas survive the encounter and escape with a reduced
velocity in either direction.  These three processes are central to the evolution
of the systems described later in this paper.  We carried out a range of calculations
to determine the conditions under which each of these outcomes can be expected.
Collisions were simulated over three initial values of the Mach number $\mach$,
the Jeans fraction per cloudlet $\fJ$, and the impact parameter $b$.  The
result of each calculation was categorised as a collapse, merger or splash, as described
above.  Finally, to ensure
that the minimum resolution requirements were met,
the calculations were repeated using 100, 300 and 1000 particles per cloudlet.

The results of the calculations are shown in table \ref{twobodytable}.  They indicate
that at low Mach numbers, the result is always a merger or a collapse.  The deciding parameter
between these outcomes is, as expected, the total number of Jeans masses, with the
calculations at $\fJ < 0.5$ always producing a merger and those at $\fJ > 0.5$
always collapsing.  In the borderline cases with $\fJ=0.5$, the more `glancing' collisions
(i.e.\ those at higher impact parameter) are less likely to collapse.
At high mach number, the result is always a splash unless the collision is head-on.

The results at the three different numerical resolutions
were unchanged except in two `borderline' cases, indicated in table \ref{twobodytable}.
In each of these calculations the result was a merger when 100 particles were used, and
a collapse at higher resolution.  This result is attributable to the number of
particles involved in the overlapping region of the collision -- at low
resolution, there are simply too few particles in the merged region to
create a sink particle.

\begin{table}
\caption{Results of two-cloudlet collisions for a range of Jean's fraction ($\fJ$),
Mach number ($\mach$) and impact parameter ($b$).  Each collision ends
in collapse ($\star$), merger ($\circ$) or splash (-).  The radial parameter $\xi$
of the relevant Bonnor-Ebert density profile is also shown.  Calculations marked \dag\
ended in collapse when $>100$ particles per cloudlet were used, and in merger
when 300 or 1000 particles were used.  The remaining calculations produced the same
outcome at all three numerical resolutions.}
\label{twobodytable}
\begin{tabular}{rllll}
$\fJ = 0.80 (\xi = 4.6)$&	$\star$&	$\star$&	$\star$&	\\
$\fJ = 0.50 (\xi = 3.0)$&	$\star$&	$\star$&	$\star$&	$b/r=0$\\
$\fJ = 0.25 (\xi = 2.1)$&	$\circ$&	$\circ$&	$\circ$&	\\
$\mach$:		&	0.5&		1&		5&		\\
			&	&		&		&		\\
$\fJ = 0.80 (\xi = 4.6)$&	$\star$&	$\star$&	-&		\\
$\fJ = 0.50 (\xi = 3.0)$&	$\star$&	$\circ$\dag&	-&		$b/r=1$\\
$\fJ = 0.25 (\xi = 2.1)$&	$\circ$&	$\circ$&	-&		\\
$\mach$:		&	0.5&		1&		5&		\\
			&	&		&		&		\\
$\fJ = 0.80 (\xi = 4.6)$&	$\star$&	$\star$&	-&		\\
$\fJ = 0.50 (\xi = 3.0)$&	$\circ$\dag&	$\circ$&	-&		$b/r=1.6$\\
$\fJ = 0.25 (\xi = 2.1)$&	$\circ$&	$\circ$&	-&		\\
$\mach$:		&	0.5&		1&		5&		\\
\end{tabular}
\end{table}

This investigation can be extended, for example to collisions between cloudlets of
unequal size and/or mass, rotating cloudlets and so forth.  However, our purpose here
is not to undertake a full analysis of two-body collisions, but
to investigate the numerical implications of the kind of collisions
that dominate the calculations presented
later in this paper.
This point is discussed further in section \ref{parameters}.

\section{Description of the system}
\label{parameters}

\subsection{General description}

The system under investigation consists of many identical, spherical
clouds of isothermal gas, which will hereafter be referred to as
`cloudlets'.  These are pressure-confined by a tenuous surrounding
medium, with which they do not otherwise interact.  Each cloudlet
initially has a stable Bonnor-Ebert density profile, and the external
pressure is chosen to match this density profile at the chosen
cloudlet mass and radius.  The cloudlets are initially uniformly
distributed in a spherical volume, and given individual bulk
velocities in random directions.  In most of the calculations the
velocities are of equal magnitudes.
The speeds of the cloudlets are then scaled to fix the virial ratio
$Q=|E\sub{K} / E\sub{G}|$ to the desired value
(where $E\sub{K}$ and $E\sub{G}$ are the overall kinetic and gravitational
energies of the cloud).

These initial conditions are not intended to be a realistic representation
of a molecular cloud.  Our calculations do not include magnetic fields, stellar feedback, heating
and cooling of the gas, or any kind of `support' mechanisms.  This allows
us to isolate the evolution of the system under the processes of gravity
and hydrodynamics. The object is to investigate a system of gas cloudlets
set up in such a way as to encourage the development of a cluster through the
processes of collisions and mergers.

In most of the calculations presented here,
the cloudlets' velocities are scaled such that $Q = 0.5$.
This is done in order to encourage the most interactions between cloudlets during the
lifetime of the cloud.  In super-virial clouds, the cloudlets will tend to fly apart
and the mean free path will increase; in sub-virial clouds, cloudlets will simply
fall to the centre.  In section \ref{dissresults} we demonstrate this
point by examining clouds with increased and decreased values of $Q$.

An isothermal equation of state has been used.
This is a good approximation to the equation of state of molecular gas
at the densities likely to be relevant in applications of the models presented here.
In molecular clouds, heating due to collisions and collapse
radiates on a short timescale, such that the gas is roughly isothermal \citep{larson69}.
This approximation holds in collapsing molecular cloud cores up to densities of $\sim 10^{13}
{\rm \; g \; cm^{-3}}$ \citep{opacitylimit}.  Hence, an isothermal equation of state
is an appropriate simplification for the simulation of gas systems such as those
presented in this paper.

\subsection{Parameters}

The properties of this system are controlled by a few key parameters:
\begin{enumerate}
\renewcommand{\theenumi}{(\arabic{enumi})}
	\item $R$, the overall radius of the cloud
	\item $N$, the number of cloudlets
	\item $\cs$, the sound speed of the gas
	\item $\fJ$, the fraction of a Jean's mass initally contained in each cloudlet
	\item $\tau$, the ratio of the collision timescale to the crossing time of the cloud ($=\tcoll/\tcross$), where
$\tcross = 2R/v$ and the cloudlets initially have a velocity $v$
	\item $Q$, the ratio of kinetic to gravitational energy.
\end{enumerate}

Since each cloudlet is a pressure-bounded Bonnor-Ebert sphere, it will obey 
\[
	\cs^2 = \alpha \frac{Gm}{r}
\]
where $m$ and $r$ are the mass and radius
of the cloudlet respectively, and
\[
	\alpha = \frac{2}{5} {\fJ}^{-\frac{2}{3}}.
\]
Hence, choosing $\fJ$ also fixes $m/r$ for each fragment.  Fixing $\tau$ is
equivalent to fixing $\lambda / R$, the ratio of the mean free path to the
radius of the cloud.  Since
\[
	\lambda = \frac{R^3}{3Nr^2}
\]
choosing $\lambda / R$ also fixes $r/R$ for each fragment.

Consequently, once the overall scaling is established by
setting $R$ and $\cs$, it is sufficient to select values of $N$, $\fJ$ and $\tau$ to
determine the mass and radius of each cloudlet.  The cloudlets must then be assigned
bulk velocities, usually with a random direction and all of the same magnitude (other
prescriptions for the initial velocity distribution are described below).  The
velocities are then scaled to
give the desired ratio $Q$ of kinetic and gravitational energy. In most models,
the cloud is virialised so that $Q=0.5$.
Finally, an important property of the system is the initial Mach number of the cloudlets
$\mach$.  Since $\mach$ indicates the typical Mach number expected
in collisions, it
is important in determining the minimum resolution, as discussed in section \ref{resolution}.

In all the calculations presented here, $N=1000$.  Table \ref{models} lists all
the models described in the text and the values of $\tau$, $\fJ$ and $Q$ chosen,
as well as the velocity distribution used.

We have simulated clouds in which all cloudlets have identical masses and radii.
This is not an attempt to model (for example) a real molecular cloud, in which a range
of `clumps' of different sizes would be seen.  Rather, the intention is to see whether
a system with a range of cloudlet sizes can arise entirely through the processes of
two-cloudlet interactions, without being specified at the outset (as suggested by
\citet{ML}, for example).

\begin{figure}
\includegraphics[width=0.45\textwidth]{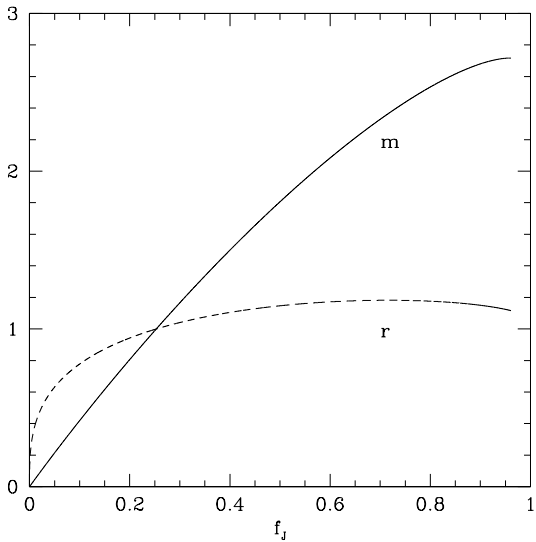}
\caption{Mass and radius of stable Bonnor-Ebert spheres over a range of Jeans fraction $\fJ$.
The temperature is fixed such that $\cs=1$, and the external pressure fixed such that
a sphere of mass $m=1$ will have a radius $r=1$.  The units are dimensionless such that $G=1$.
Note that there is a maximum sphere radius, which occurs when $\cs^2 = Gm/2r$.}
\label{sizerange}
\end{figure}

Furthermore, the possibility of setting up a range of cloudlet sizes in one cloud
is limited by the restriction that the cloudlets are stable Bonnor-Ebert spheres.
This is indicated in figure \ref{sizerange}, which shows the mass and radius of
stable Bonnor-Ebert spheres at a range of $\fJ$, subject to a fixed temperature and
external pressure.  The stable solutions all have roughly the same radius, except at very
low $\fJ$ (at which values, a very large number of mergers would be required to
form a gravitationally unstable object),
and so it would not be feasible to set up a cloud in which the cloudlets had
a significant range of radii.  Models with high values of $\fJ$ are described in section
\ref{protresults}.

\begin{table}
\caption{Definitions of the parameters used in the various models referred to in the text.}
\label{models}
\begin{tabular}{c...c}
Model	&	\tabc{$\tau$}	&	\tabc{$\fJ$}	&	\tabc{$Q$}	&	Velocities	\\
1	&	1		&	0.25		&	0.5		&	i		\\
2	&	0.2		&	0.25		&	0.5		&	i		\\
3	&	5		&	0.25		&	0.5		&	i		\\
4	&	1		&	0.25		&	0.5		&	d		\\
5	&	1		&	0.25		&	0.01		&	i		\\
6	&	1		&	0.2		&	1.0		&	i		\\
7	&	1		&	0.5		&	0.5		&	i		\\
8	&	1		&	0.8		&	0.5		&	i		\\
9	&	1		&	0.25		&	0.5		&	e		\\
10	&	1		&	0.25		&	0.5		&	c		\\
\end{tabular}

\medskip
The symbols $\tau = \tcoll/\tcross$, $\fJ = m\sub{cloudlet}/M\sub{J}$ and
$Q = | E\sub{K} / E\sub{G} |$ are defined in the text.  The letters in the Velocities
column refer to the initial velocity dispersion and are as follows:
(i) isotropic and uniform, (d) isotropic with a distribution of speeds,
(e) elliptically distributed and (c) spatially correlated.
\end{table}

\section{Calculation results}
\label{results}

The `baseline' calculation (model 1) used $N=1000$ cloudlets,
$\fJ=0.25$ and $\tau=1$.
Under these conditions, the cloud initially contains 250 Jean's masses and the
collision timescale $\tcoll$ is equal to the crossing time $\tcross$.
The virialisation of the cloud results in each cloudlet being given a bulk
velocity of $\mach \simeq 2.7$.  Consequently, random collisions between cloudlets
can be expected to take place at Mach numbers up to around $\mach=5$.
Each cloudlet contained 100 SPH particles.
Several random realisations of
the same system were run to ensure that the result was independent of the
specific initial conditions.

\begin{figure}
\includegraphics[width=0.22\textwidth]{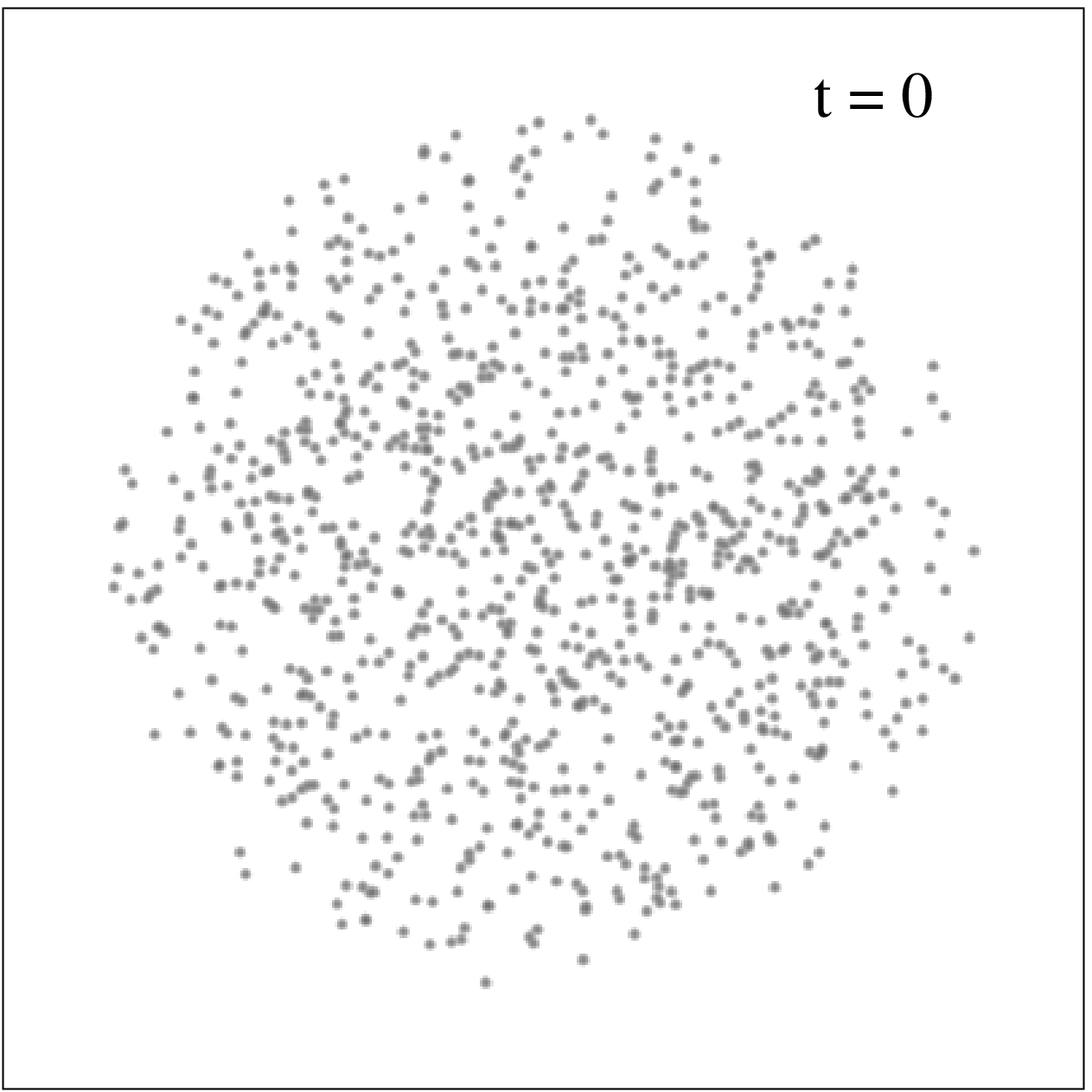}
\includegraphics[width=0.22\textwidth]{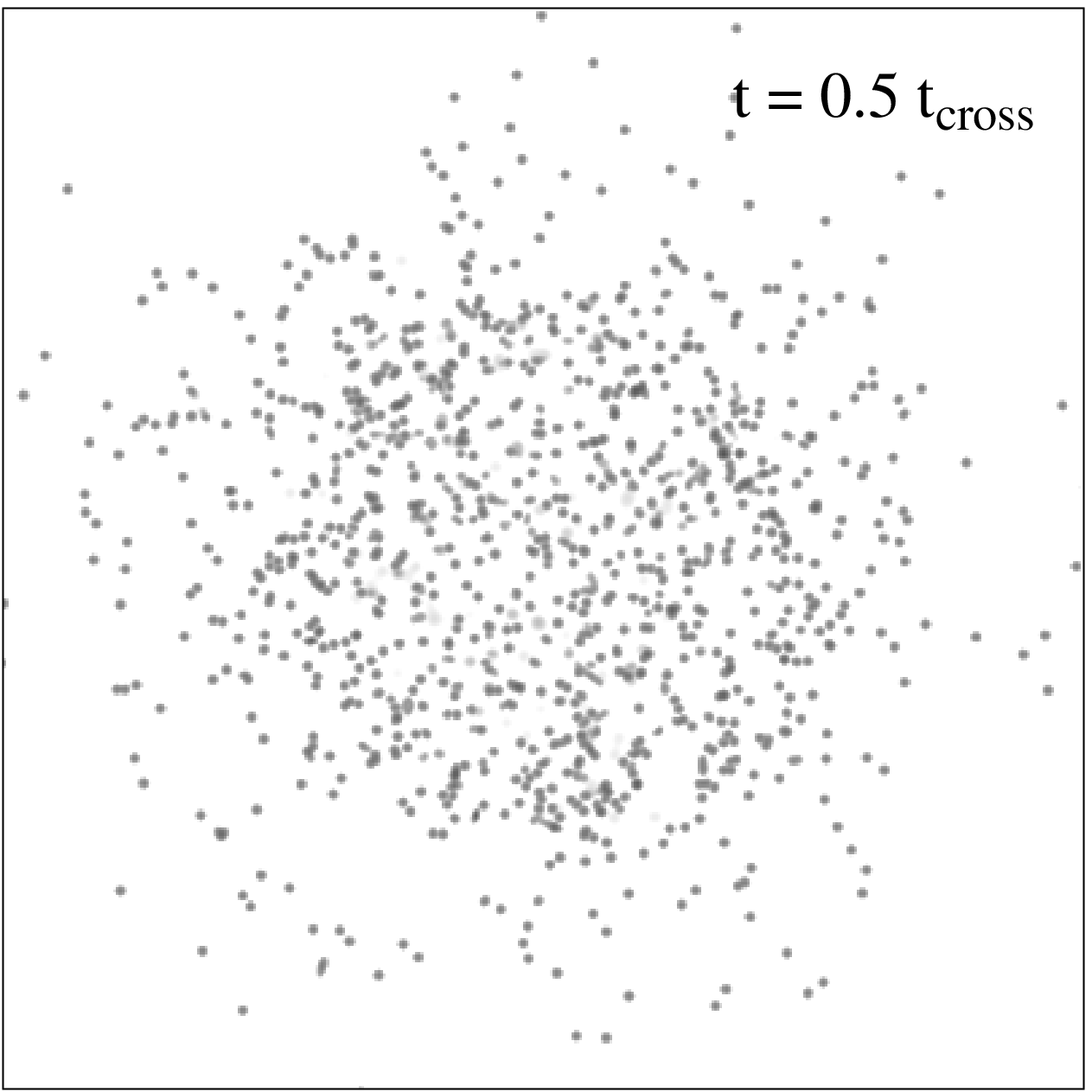}
\includegraphics[width=0.22\textwidth]{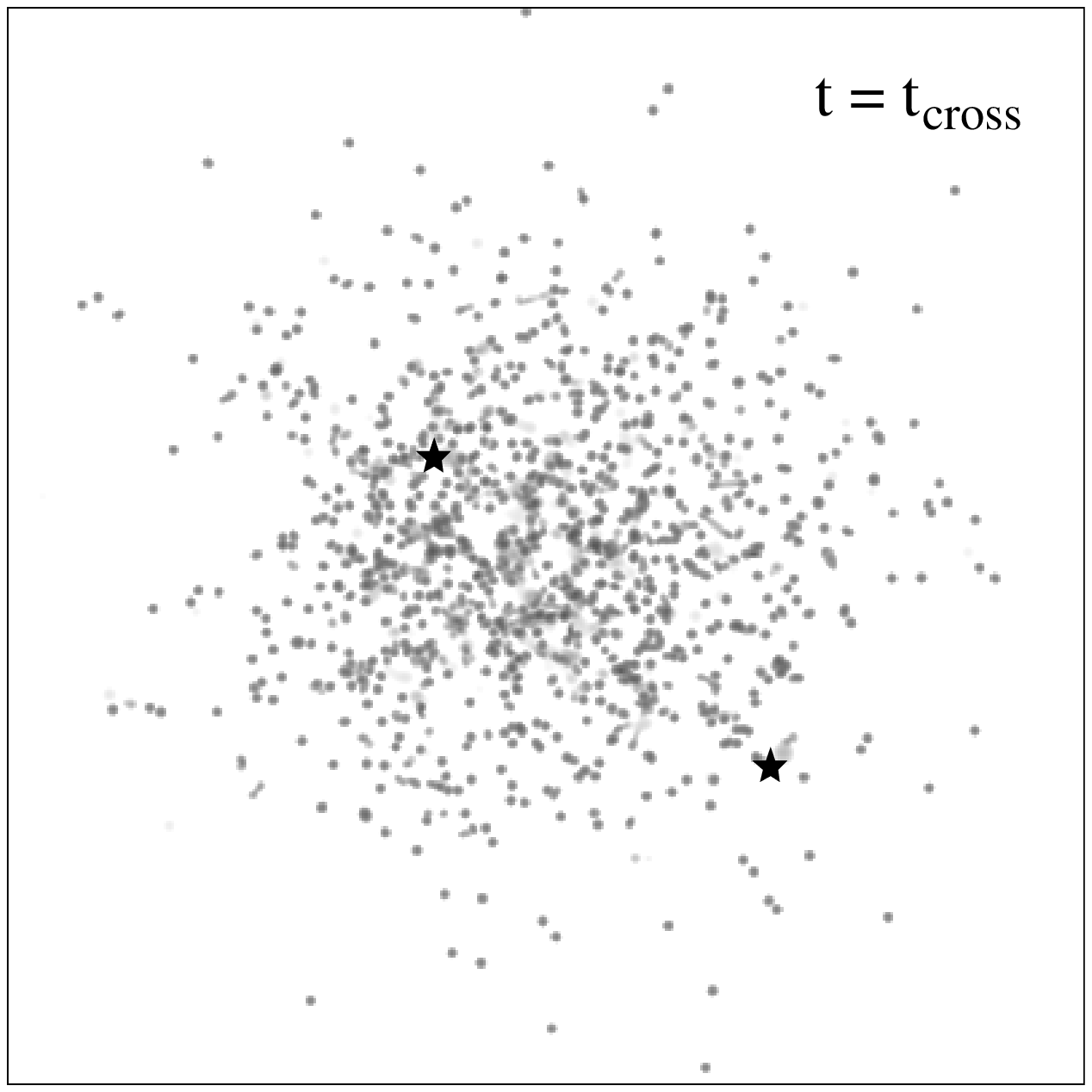}
\includegraphics[width=0.22\textwidth]{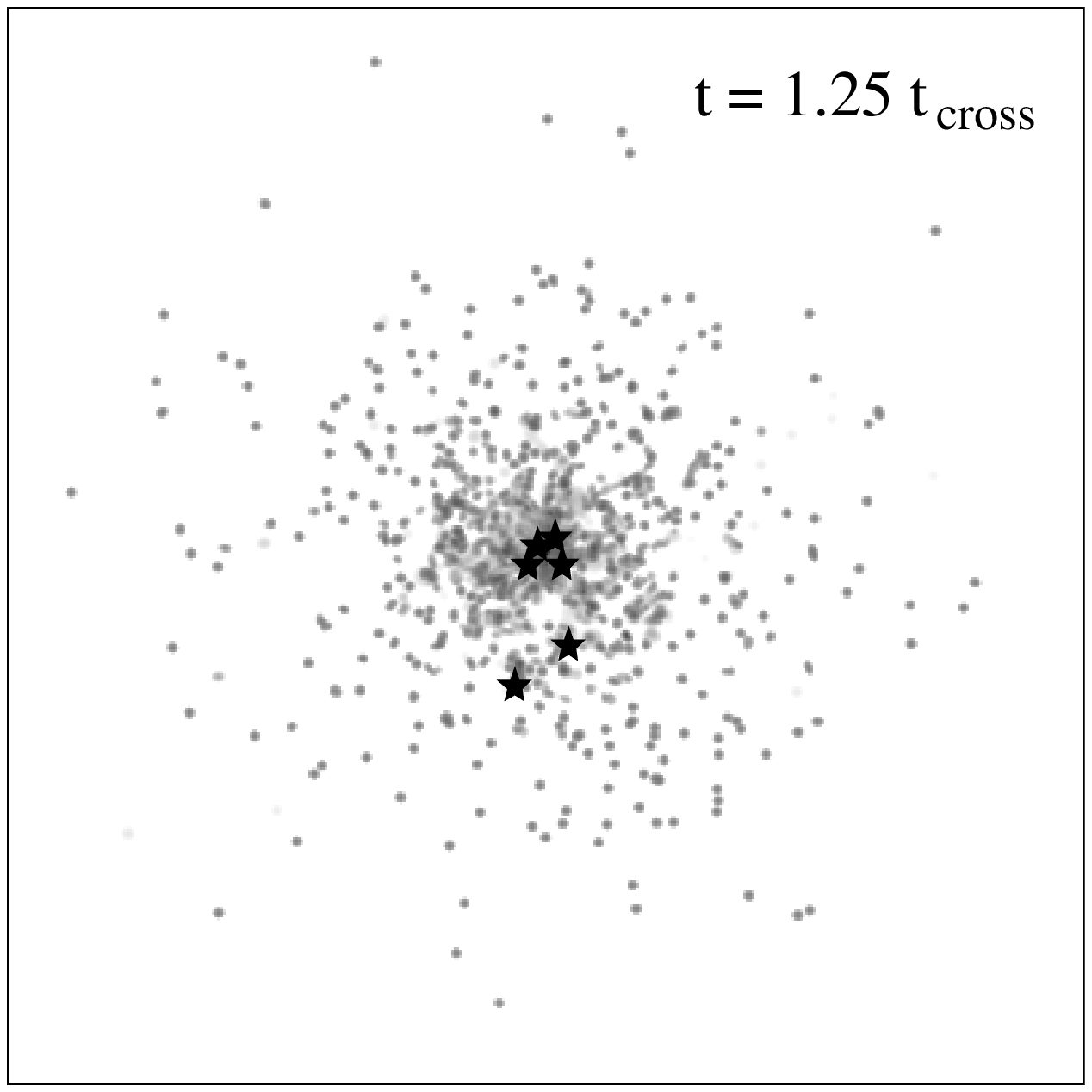}
\caption{Column-density maps of the cloud used in model 1, at times $t=0$, $t=0.5\tcross$,
$t=\tcross$ and $t=1.25\tcross$.  Protostars are indicated by star symbols.}
\label{A1pics}
\end{figure}

The general evolution of the system is shown in figure \ref{A1pics}.  As the cloudlets collide, their
relative kinetic energy is dissipated, and material is `splashed' into the region
between them as they separate.  This material falls toward the centre of the gravitational
well.  After a time $\tcoll$, most of the
gas has formed a large dense region in the centre of the cloud.  The gas in this
region continues to collapse, and goes on to form a complex structure in which
locally gravitationally unstable regions form.  However, these processes are not
resolved in our calculations, and consequently do not bear close examination.
Such complex turbulent regions must currently be studied separately using very
high resolution (see for example \citealt*{bate02}).

The calculation consists of two phases.  To begin with, kinetic energy is
dissipated approximately linearly with time as the cloudlets collide with one another.  After each
collision, splashed material, which tends to lose specific angular momentum in the collision,
begins to fall to the centre.
This continues until roughly $t=\tcoll$, at which point the average number of collisions
per cloudlet is expected to be about 1.  By this time, a significant fraction of
the gas has fallen to the centre of the cloud.  Cloudlets passing through this central
region are disrupted and add to the gas already present.  The system enters a second
phase, in which the kinetic energy is rapidly dissipated, and the dense gas at the centre
begins to produce local gravitational collapses as described above.

Within this dense central region, complex
interactions of cloudlets (involving more than two cloudlets at a time)
and disrupted material will occur.  However, an examination
of such processes is not the intent of this paper, and (as mentioned above)
these regions are not resolved in our calculations.  We will not, therefore,
discuss here the further fate of material that has fallen into a
region of runaway collapse, leaving such work to more detailed
numerical investigations.

Given this fact, the evolution of the system is dominated by two-body
collisions.  Interactions between more than two cloudlets (outside
the central region) are rare.

\subsection{Dissipation timescale results}
\label{dissresults}

\begin{figure}
\includegraphics[width=0.48\textwidth]{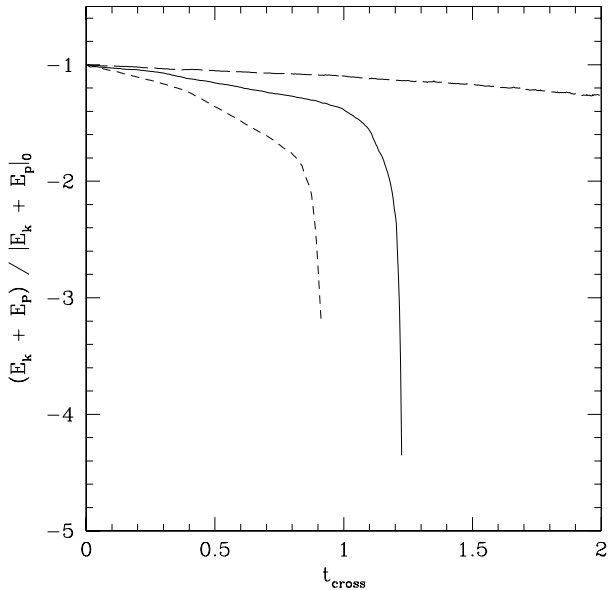}
\caption{The decay of total energy vs.\ time in model 1 ($\tau = 1$, solid),
model 2 ($\tau = 0.2$, short dashed) and model 3 ($\tau=5$, long dashed).}
\label{A1diss}
\end{figure}

It is of particular interest to examine the timescale on which the evolution of
the system takes place.  Figure \ref{A1diss} shows the change in the total
(e.g.\ kinetic plus potential) energy of the system
(in which the collision time $\tcoll \simeq \tcross$) with time.
It is clear that in this case, the kinetic energy is dissipated on the collision timescale, $\tcoll$.
Also shown in figure \ref{A1diss} are the results from two further calculations
with $\tau = 0.2$ and $\tau = 5$ (models 2 and 3 respectively), in order to clearly indicate the
dependence of the energy dissipation on $\tcoll$.

When $\tcoll$ is
reduced by a factor of five, the initial rate of dissipation of kinetic energy is approximately
five times greater
as expected.  The runaway dissipation begins at a time $t \simeq 0.8 \tcross$.
This broadly agrees with the simple expectation that cloudlets will begin to fall
to the centre after a time of roughly $\tcoll$ ($=0.2\tcross$ in this case),
and take around half a crossing time to do so.
Similarly, when $\tcoll$ is increased by a factor of five, the initial
dissipation rate is reduced by roughly the same factor.  In this case, the
runaway collapse phase is not reached by $t = 2 \tcross$.

Scalo \& Pumphrey (1982) performed calculations of a similar system of cloudlets,
using a modified N-body code.  The parameters
of the system used in their calculation are essentially the same as that used in ours,
except that they did not include gravity -- the cloud was kept confined by using a hard
reflective outer boundary.
They concluded that the dissipation time in such a system is at least an order of magnitude
longer than the collision time, due to the propensity of collisions at low impact parameter.

In our calculations, the system will not survive for more than roughly the collision time.
There are two main reasons for this difference.
The first is the `splashing' of material in collisions in our calculations.  The simulations
of Scalo \& Pumphrey used prescribed rules in which the outcome of any collision was either one, two
or three discrete cloudlets, at the same density as the colliding clouds.  Consequently,
there were no spread-out regions of low density gas, which would otherwise have
added to the total collisional cross-section
of the gas as the calculation proceeded.  Secondly, since
no gravity was included, the cloud did not relax toward a central density profile,
and there was no preferred centre toward which gas would fall after a collision.  Hence
the formation of a central dense region, causing the runaway dissipation in the second
phase of our calculations, was impossible.

We also examined the effect of introducing a distribution of initial cloudlet
speeds, by repeating the calculation as follows (model 4).
Cloudlets were given velocities in a random direction, and magnitudes
chosen randomly from a probability distribution $p(v)$, with
\[
	p(v) \propto v^2 \exp(-v^2).
\]
The velocities were then scaled as before to ensure overall viriality.  All other
parameters of the calculation were unchanged.  In this case,
the initial rate of energy dissipation was equal to that seen in model 1,
but the cloud went into runaway collapse after around $0.5\tcross$ (compared to
model 1 which took twice as long to reach this phase).  This difference is
attributable to the cloudlets initially at low speeds, which tend to simply
fall to the centre of the cloud and arrive at the same time, creating
a dense region and initiating the collapse.

In two further runs, we examined the effect of altering the virial coefficient
$Q = \left| E\sub{K} / E\sub{G} \right|$ (for all models except 5 and 6, $Q=0.5$).
Model 5 is highly sub-virial, such that $Q=0.01$,
and is otherwise identical to model 1.
In model 6 the cloud is super-virial with $Q=1$, and in order to satisfy resolution
requirements, the Jeans fraction has been reduced to $\fJ=0.2$.  In these calculations
the crossing time (as defined from the initial velocity of the cloudlets) is not very meaningful,
so instead we will use the gravitational free-fall time, given by
\[
	t\sub{ff} = \left( \frac{\pi^2 R^3}{8GM} \right)^{\frac{1}{2}}.
\]
This is roughly half the crossing time $t\sub{cross} = 2R / v$ for virialised clouds, confirmed by setting
$v^2 = GM/R$ in the above to recover
\[
	\frac{t\sub{ff,virial}}{t\sub{cross}} = \frac{\pi}{2\sqrt{8}} \simeq 0.56.
\]

\begin{figure}
\includegraphics[width=0.48\textwidth]{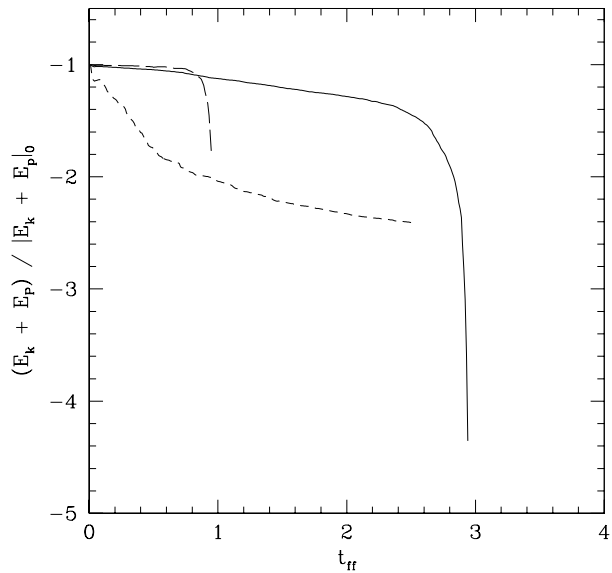}
\caption{The decay of total energy vs.\ time in model 1 ($Q=0.5$, solid),
model 5 ($Q=0.01$, long dashed) and model 6 ($Q=1$, short dashed).}
\label{hotcold}
\end{figure}

The change in total energy in models 5 and 6 compared to model 1 is shown
in figure \ref{hotcold}.  The results are largely what is expected.  In
model 5, the cloudlets simply fall to the centre.  Some energy dissipation occurs
as they merge, but the dissipation rate is somewhat reduced
compared to the virialised cloud.
Several protostars are formed through these mergers, although
by the time the first of these has formed, the maximum density has increased beyond the
limit set by the resolution criterion set out in section \ref{resolution}, and
consequently the gravitational collapse at all of these points is not guaranteed
to be realistic.
After about one free-fall time,
the cloud goes into runaway collapse at the centre.  In model 6, however, the clouds
undergo high Mach number collisions from the outset and the initial rate of energy dissipation
is high.  This rate gradually decreases as the cloud increases in size and the mean free path
correspondingly grows.

\subsection{Protostar distribution results}
\label{protresults}

An important question to ask of the system under investigation is whether it
can lead to a cluster of protostars, and if so, what the spatial distribution
of those protostars will be.  These `protostars' are represented in our
numerical code by replacing collapsing regions of gas with `sink particles';
see section \ref{code} for a description of the formation and accretion
criteria used.

In the baseline calculation, the resulting distribution of protostars was highly
clustered in the centre of the cloud.  All of these were formed by the fragmentation
of the large mass at the centre - none were formed in two-cloudlet interactions.
This result holds independently of the collision timescale of the cloud.

Since the Jean's fraction in each cloudlet, $\fJ$, is only $0.25$, four cloudlets
need to be assembled together before one Jean's mass can be reached.  If, however
$\fJ$ is increased, a smaller number of cloudlets will be required, and the formation
of protostars in cloudlet interactions should become more likely.  Hence, two further
calculations were performed, with $\fJ=0.5$ (model 7) and $\fJ=0.8$ (model 8),
in order to encourage
as far as possible the formation of protostars throughout the volume of the cloud
rather than in the centre only.

\begin{figure}
\includegraphics[width=0.22\textwidth]{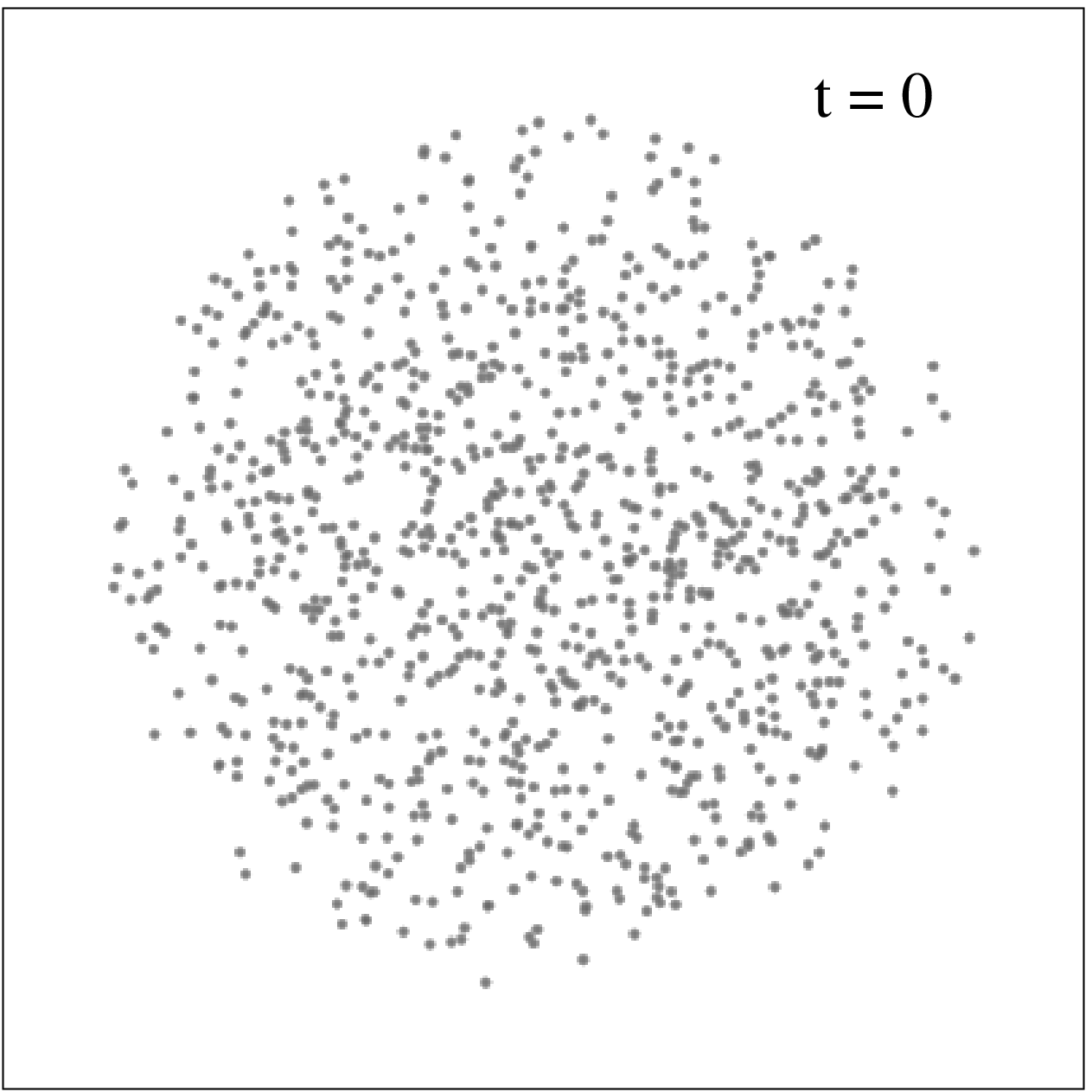}
\includegraphics[width=0.22\textwidth]{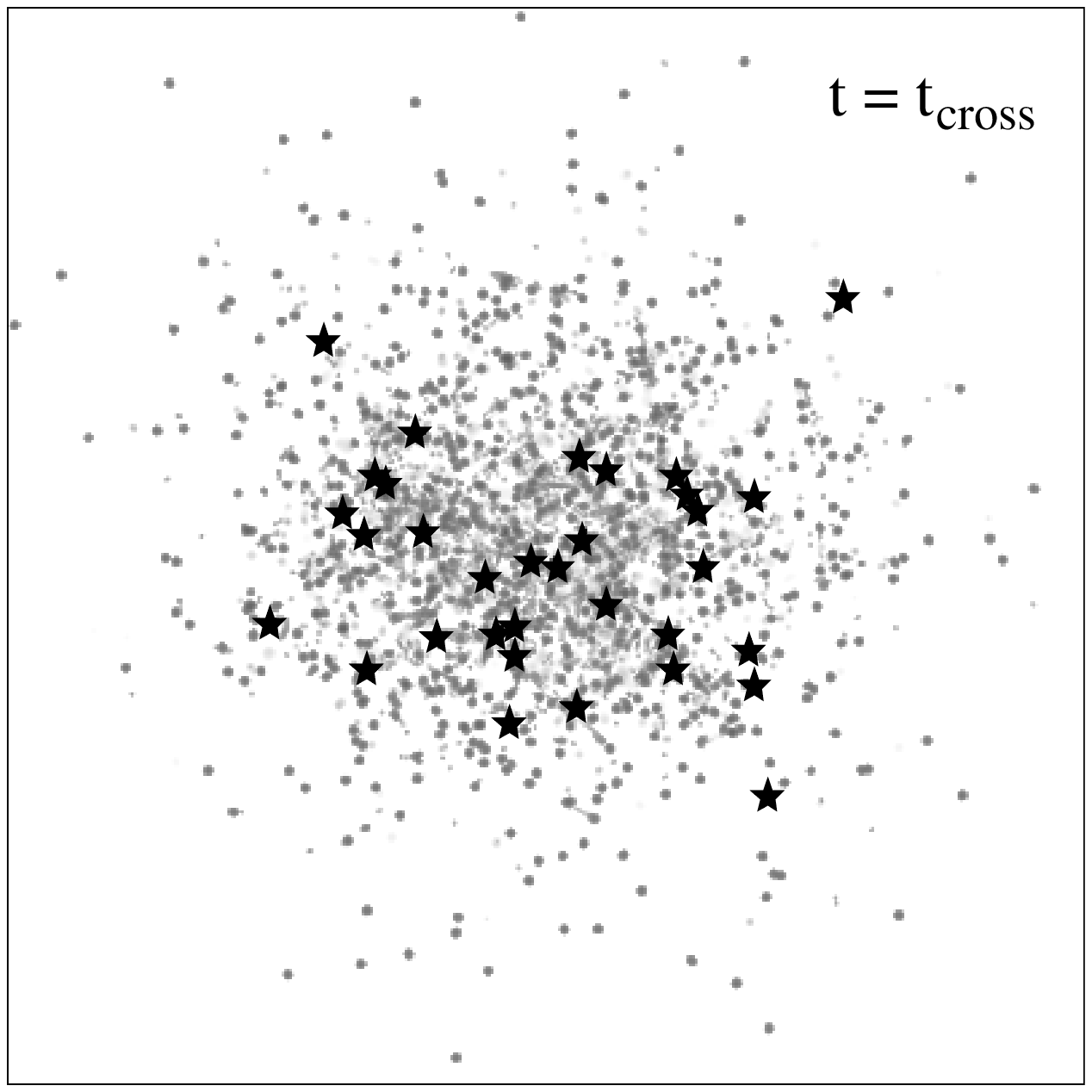}
\caption{Column-density maps of model 7 ($\fJ = 0.5$) at time $t=0$ (left) and $t=\tcross$ (right).
Protostars are indicated by star symbols.}
\label{fJ0.5pics}
\end{figure}

\begin{figure}
\includegraphics[width=0.22\textwidth]{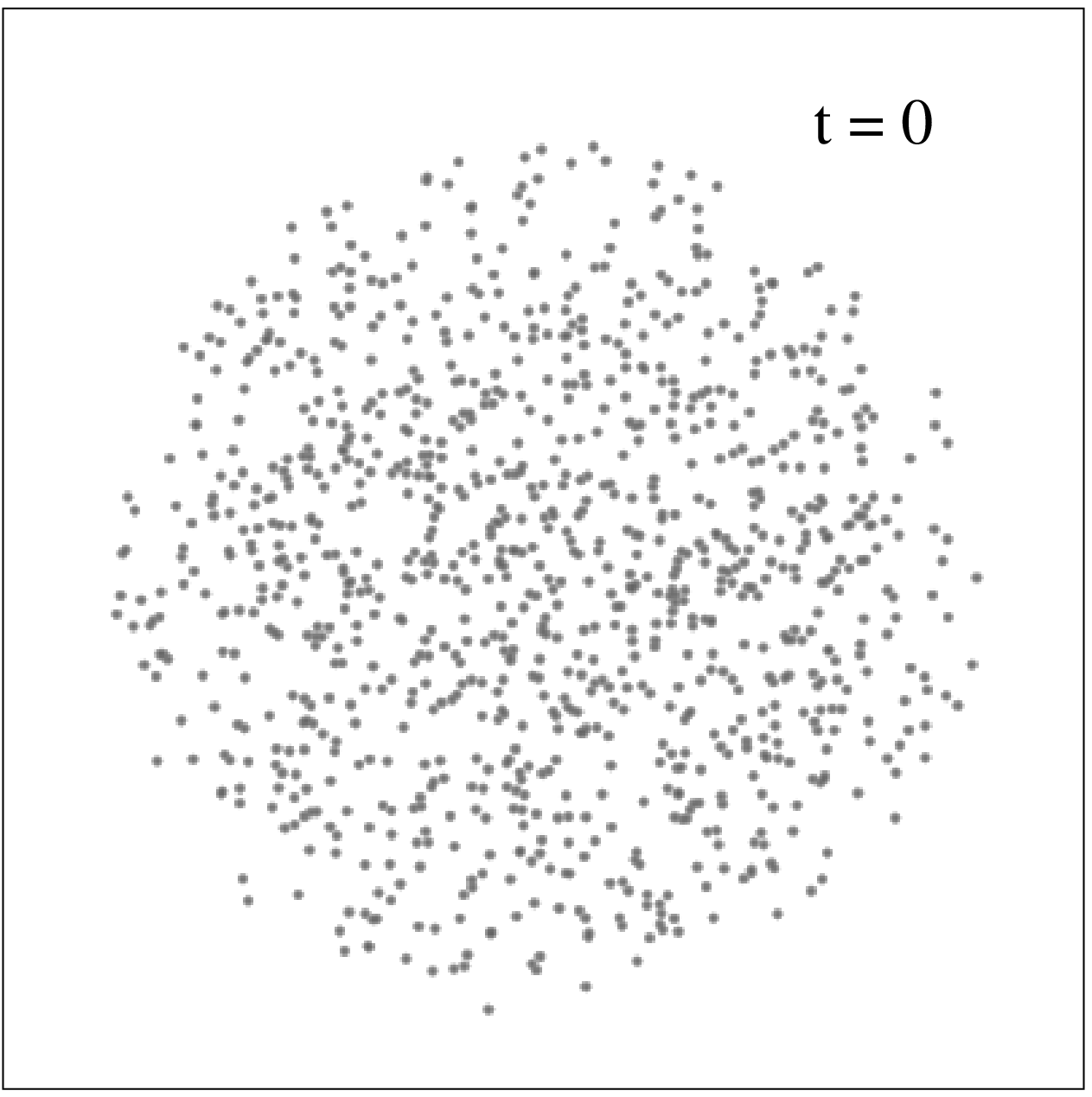}
\includegraphics[width=0.22\textwidth]{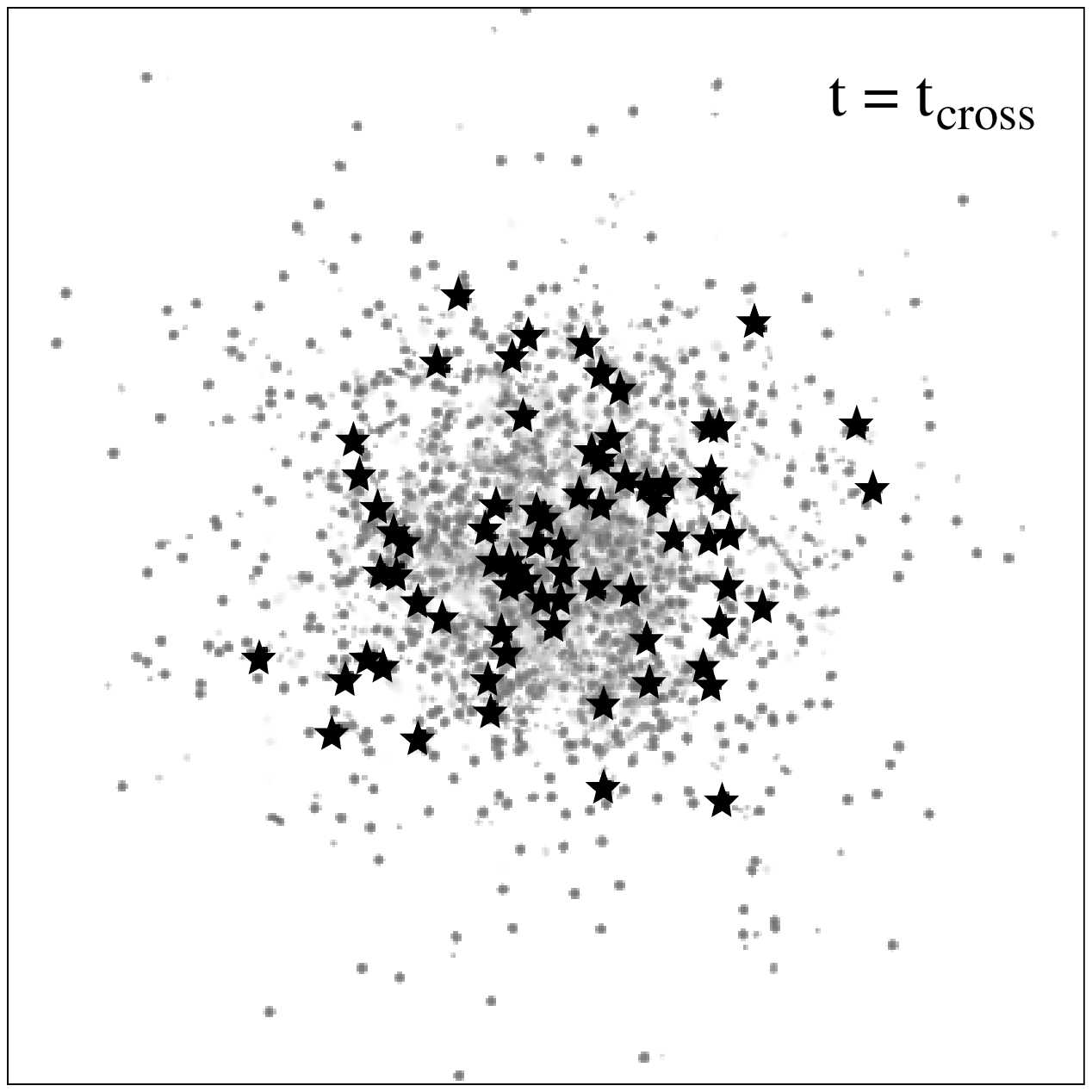}
\caption{Column-density maps of model 8 ($\fJ = 0.8$) at time $t=0$ (left) and $t=\tcross$ (right).
Protostars are indicated by star symbols.}
\label{fJ0.8pics}
\end{figure}

The results of these calculations are shown in figures \ref{fJ0.5pics} and \ref{fJ0.8pics}.  Protostars are
now formed in all parts of the cloud in two-cloudlet collisions, and as expected,
more protostars are formed when $\fJ = 0.8$ than when $\fJ = 0.5$.  However,
at the end of the calculation, only a small proportion of the initial mass of the cloud
(0.1 for model 7 and 0.15 for model 8)
is present in these protostars -- the remainder once again forms a large mass
at the centre of the cloud.  Thus, although increasing $\fJ$ significantly does
result in the formation of some protostars throughout the cloud, the majority
of the mass still falls into a small region in the centre.

\begin{figure}
\includegraphics[width=0.22\textwidth]{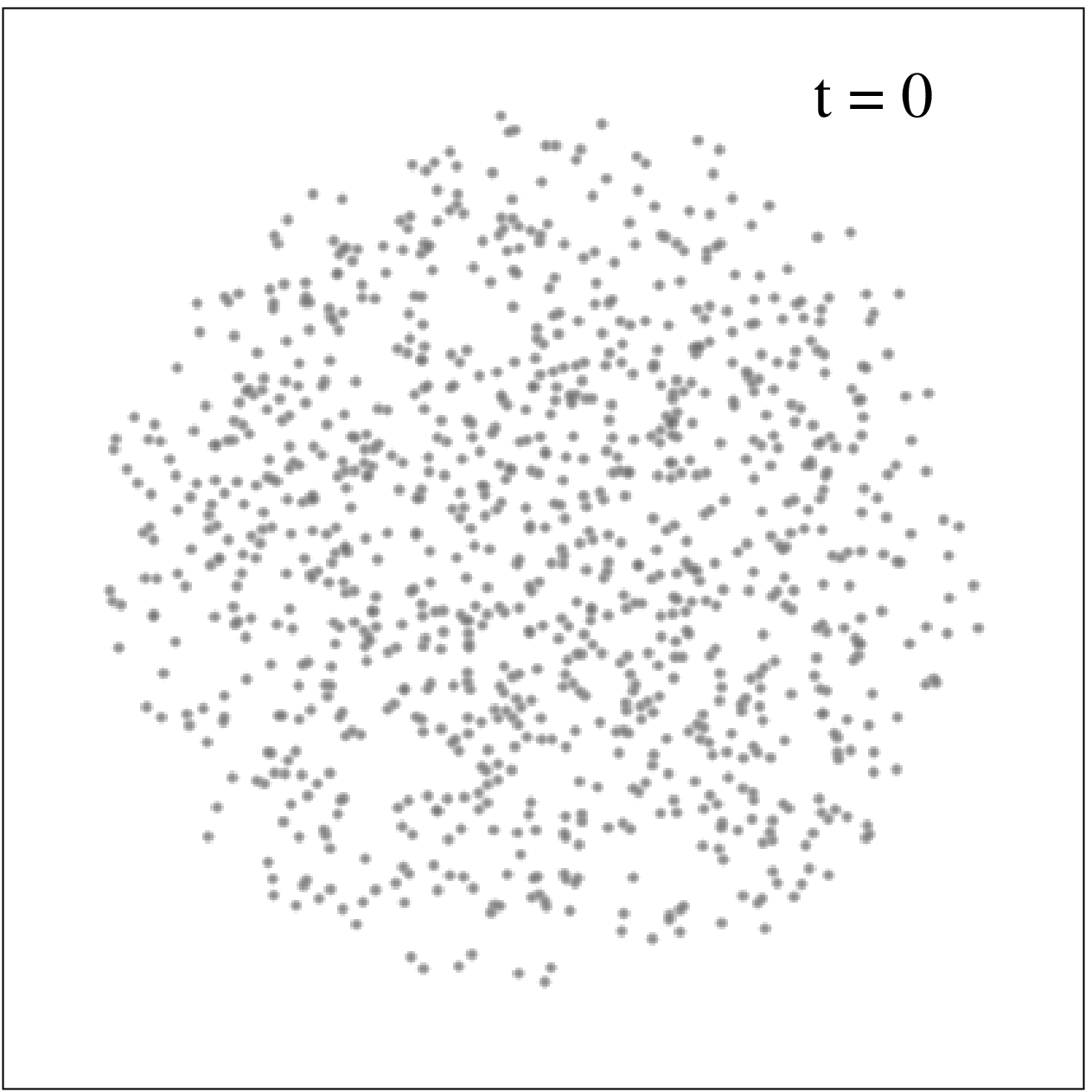}
\includegraphics[width=0.22\textwidth]{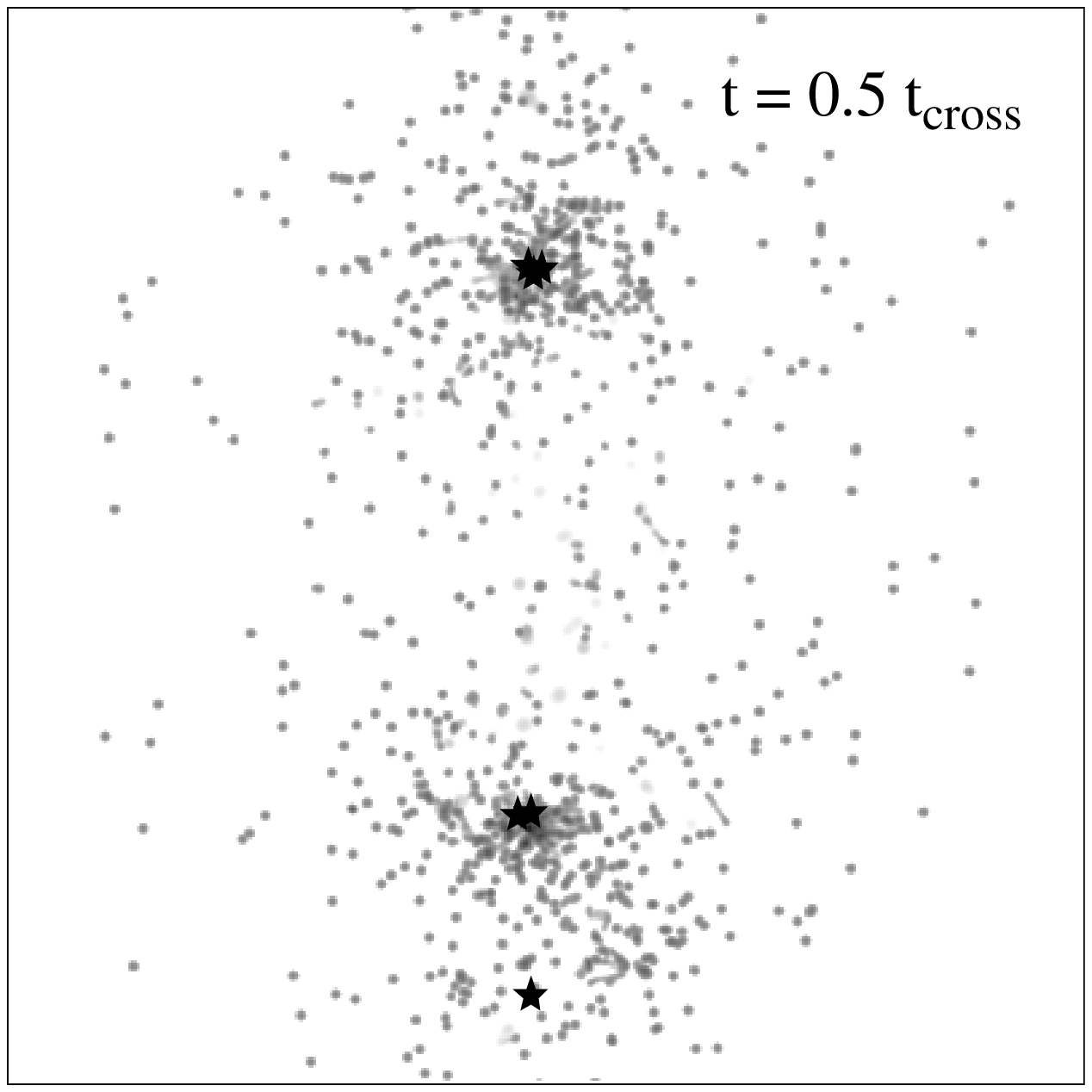}
\caption{Column-density maps of model 9 at time $t=0$ (left) and $t=0.5\tcross$ (right).
Protostars are indicated by star symbols.}
\label{ellipticalpics}
\end{figure}

The central distribution of the mass in the final state of these
calculations is a consequence of the isotropic velocity distribution.  A more
complicated initial velocity distribution might be expected to result in
a corresponding change in the eventual outcome.  In order to investigate
this, further calculations were performed, in which the velocity distribution
took the form of a prolate ellipsoid, such that the cloudlets' velocities
were preferentially aligned along one axis (model 9).
The results of these calculations
are shown in figure \ref{ellipticalpics}.

The effect of the velocity anisotropy is clear: rather than forming one
large central mass, the gas has fallen into two dense regions at the foci
of the velocity ellipsoid.  Otherwise, the results are largely unchanged; the
kinetic energy of the system is still dissipated in roughly $0.5 \tcoll$,
and local gravitational collapse is still confined to the two dense
regions.

\subsection{Coagulation}

It is known from the previous results outlined in section \ref{mergers} that cloudlets
encountering each other at a sufficiently low Mach number (which is around 1)
will merge to form a larger cloudlet.  If a sufficient number of cloudlets merge
together, at some point they will surpass the gravitational stability limit
and start to collapse.  If this process continues throughout the evolution of the cloud, the
continual mergers will lead to a distribution of cloudlet mass, and many collapsed
objects could be formed by this mechanism.

This idea has been discussed as an origin for the IMF
in globular clusters by \citet{ML}.
They examined somewhat similar systems to those discussed in this paper,
using an N-body code in which cloudlets instantaneously merge when they touch.
They found that this process resulted in a mass power spectrum of protostellar objects
at the end of the calculation.
It might, therefore, be expected that a similar process will occur in our
simulations, leading to a power spectrum of final collapsed objects.

In fact, very few mergers occur and the coagulation process is largely
unimportant in our calculations.  The main reason for this is simply that
the great majority of collisions occur at too high a Mach number.  The
prescribed rule that any two cloudlets will merge on contact, used by Murray \&
Lin, is replaced in our work by the hydrodynamical result of the collision,
which is (in general) a dissipative `splash' rather than a merger.

The typical Mach number of collisions in the system does, of course, depend
on the initial parameters.  If the initial velocity distribution is isotropic
and the velocities are scaled such that the cloud is virialised, then the
relation between the mach number and the remaining parameters $N$,
$\tau \cong \lambda/2R$ and $\alpha$ can be found.  Starting from
\[
	\mach = \frac{v}{\cs}
\]
and using $v^2 \sim M/R$, $\cs^2 \sim \alpha m/r$ and $M=Nm$, we have
\[
	\mach^2 \sim \frac{Nr}{\alpha R}.
\]
Then, since
\[
	\tau \sim \frac{1}{N} \left( \frac{R}{r} \right)^2
\]
we find
\[
	\mach^4 \sim \frac{N}{\tau \alpha^2}.
\]

Using the parameters $\alpha = 1$, $N=1000$ and $\tau=1$ the initial velocity
of the clumps is around Mach $2.5$, leading to collisions at up to
Mach 5.  If we wish to ensure that all collisions are subsonic, the initial velocities
need to fall by a factor 5.  For fixed $N$ and $\alpha$, this implies 
$\tau=625$.  At this value, a given clump will be expected to have only
one collision in 625 crossing times.  Alternatively, if $\tau$ is fixed then
either (for fixed $N$) $\alpha = 25$, which corresponds to $\fJ \simeq
2 \times 10^{-3}$, or (for fixed $\alpha$) $N = 1.6$.  The latter case is
clearly not interesting, and in the former, around half of all the cloudlets would
need to merge before one Jeans mass could be reached.

The conclusion is that the typical Mach number of collisions in the system, if
the cloud is virialised and the velocity distribution is uniform,
cannot be reduced to subsonic level without resorting to very long mean free paths,
very stable initial cloudlets, or both.

It is possible to encourage collisions at lower velocities by introducing
a spatially correlated initial velocity field.  The Larson relations \citep{larson81}
state that the velocity dispersion in star-forming clouds varies with the length
scale $l$ as $l^\sigma$, where $\sigma \simeq 0.38$.  This
correlation can be applied to an ensemble of cloudlets by
repeatedly subdividing the volume of the calculation into eight cubes, then subdividing
each cube again, and at each level assigning each cube random
velocities on the appropriate scale.  Murray \& Lin examined systems both
with and without this property.

In the case of a cloud of 1000 cloudlets, subdividing the volume three times
leaves only one or two cloudlets remaining in each subcube.  Thus, the length scale
at the smallest subdivision will be a factor $8$ smaller than the overall cloud.
The velocity dispersion on this scale will consquently be reduced by a factor
$8^{0.38} \simeq 2.2$ compared to the overall cloud.  This can be expected
to produce many more collisions at low Mach number, and hence promote
the formation of protostars due to agglomeration of several cloudlets.

\begin{figure}
\includegraphics[width=0.22\textwidth]{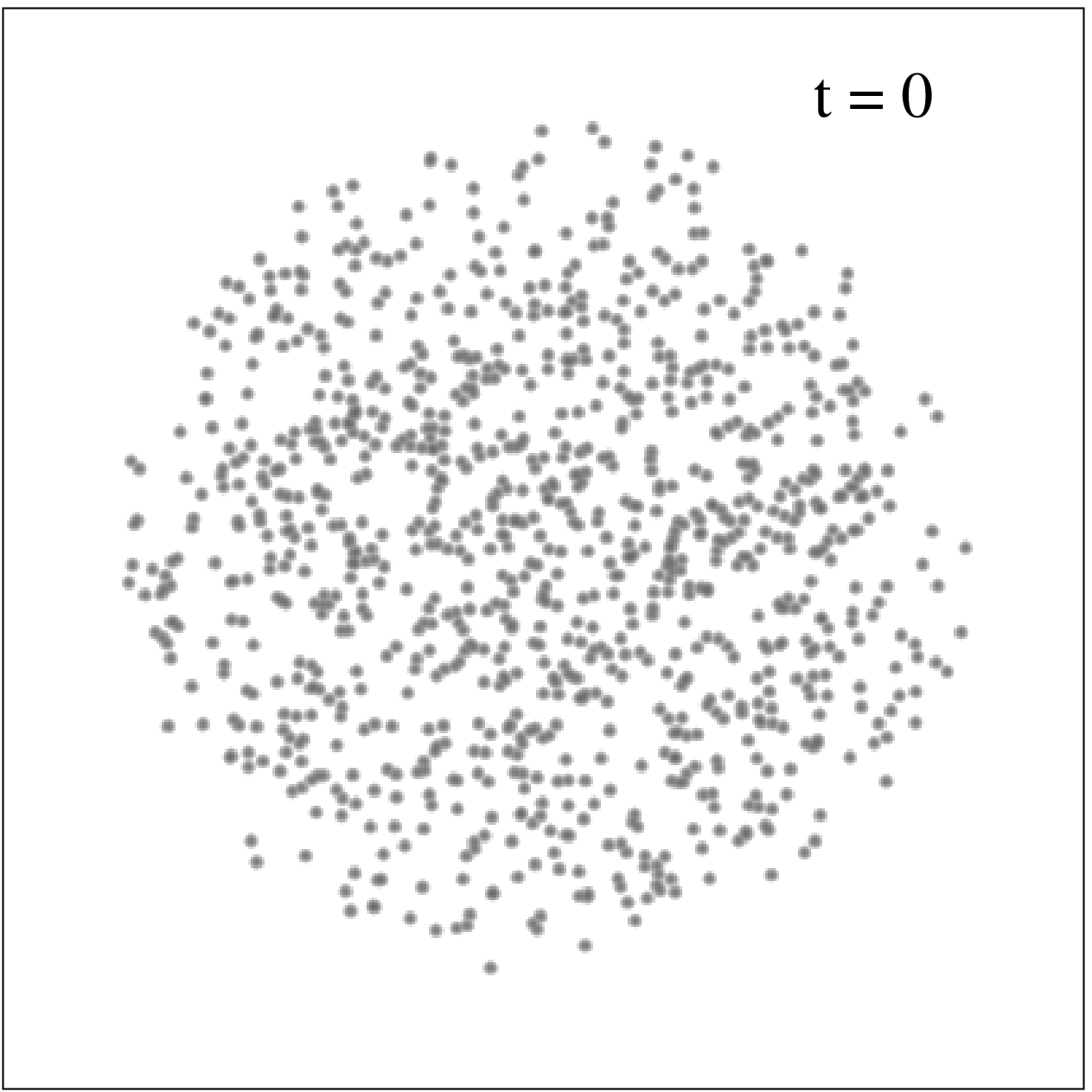}
\includegraphics[width=0.22\textwidth]{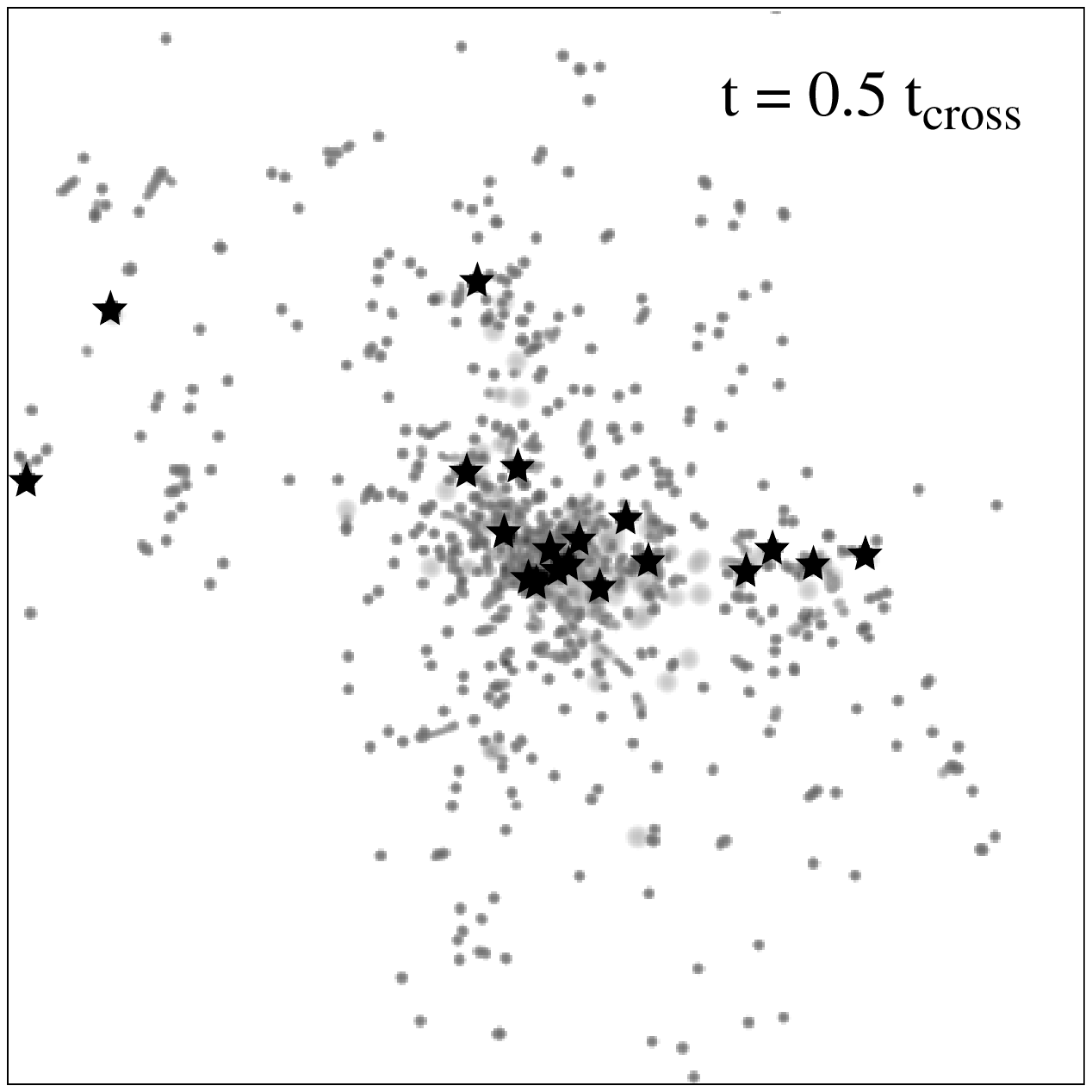}
\caption{Column-density maps of model 10 at time $t=0$ (left) and $t=0.5\tcross$ (right).
Protostars are indicated by star symbols.}
\label{correlatedpics}
\end{figure}

To this end, we repeated the calculation of model 1, with the introduction
of a spatially correlated velocity field (model 10).  The cloud is repeatedly
subdivided as described above, and the velocity for each subcube is chosen
with a random direction and a magnitude picked from a gaussian of the
appropriate width.  The velocities are once again scaled to ensure overall viriality.
The results are shown in figure \ref{correlatedpics}.

In this calculation, mergers did occur, and several protostars were formed
outside the core of the cloud by the collapse of merged cloudlets.  However,
only a small number of protostars were formed before the system once again
formed a runaway collapse region at the centre.  This occurred at $t \simeq 0.5 \tcross$,
as is expected, since the cloudlets in this model begin with a spread of initial
speeds similiar to that used in model 4 (see section \ref{dissresults}).

\begin{figure}
\includegraphics[width=0.45\textwidth]{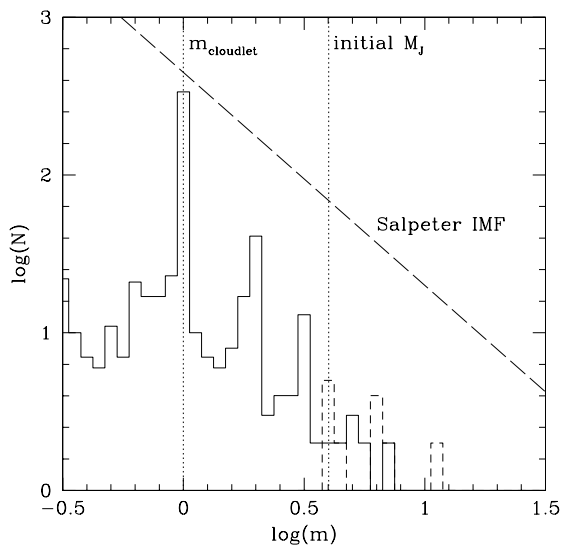}
\caption{Mass distribution of uncollapsed objects (solid line) and protostars
(dashed line) from model 10 at $t=0.5\tcross$.  The vertical dotted lines indicate the
initial cloudlet mass and initial Jeans mass.  The slope of the Salpeter IMF is indicated
by the long dashed line.}
\label{massdist}
\end{figure}

The mass distribution of cloudlets at this time is shown in figure \ref{massdist}.
The distribution of protostar masses is also shown, but note that the majority of
these formed in the unresolved cloud centre with masses above $m=10$ and are not
shown.
The peak at $\log(m)=0$ represents undisturbed cloudlets with $m=1$, while
the following two peaks correspond to $m=2$ and $m=3$, indicating the number
of complete two-cloudlet and three-cloudlet mergers.  Many cloudlets with masses
below $m=1$ are formed from material sheared off in collisions, and these are in fact
more numerous than cloudlets formed at masses $m>2$ from mergers, reinforcing the point
that disruptive collisions play a more important r\^{o}le in the system than mergers.
The slope of the Salpeter IMF \citep{salpeter} is also shown, and it is clear that
the distribution of merged cloudlets is too steep to be consistent with a stellar IMF.

\section{Conclusions}
\label{conclusion}

The general evolution of a system of cloudlets as laid out in this paper
tends to consist of two phases: an initial dissipative phase in which
kinetic energy is lost at a (roughly linear) rate determined by
the collision timescale, and a second collapse phase in which the 
majority of the gas falls to dense regions and becomes gravitationally unstable.
This second phase will begin much sooner if a broad distribution of initial cloudlet
speeds (rather than a $\delta$-function) is used.
The evolution timescale of the whole cloud is thus affected by the mean
free path of the cloudlets, and the initial velocity distribution
also plays a significant part in determining the cloud lifetime,
which can be significantly different from the dynamical timescale of
the cloud.

Protostar formation is mostly confined to the dense region that forms
in the centre of the cloud, even if the cloudlets are initially close
to gravitational instability.

Finally, we find that the process of hierarchical assembly of cloudlets
through mergers to form a mass spectrum of final objects is not significant.
This is because the great majority of collisions occur at too high a Mach
number and do not result in merging of the cloudlets.  This will remain
true so long as: the collision timescale is comparable to the crossing time
of the cloud, and the cloudlets are not of such a low mass as to be
very far from gravitational instability.  Introducing a spatially
correlated velocity field can promote mergers, but the effect is
still small and few protostars will be formed by this mechanism.

\section{Acknowledgments}
Some of the computations reported here were performed using the UK Astrophysical Fluids Facility (UKAFF).
We thank Prof.\ Jim Pringle and Dr.\ Ian Bonnell for their input and interest in this work.
We also thank the anonymous referee for their constructive comments and suggested improvements.

\end{document}